\documentclass[draftclsnofoot,onecolumn,12pt]{IEEEtran}

\usepackage{lipsum}

\usepackage[tbtags]{amsmath}
\usepackage{amsmath}

\usepackage{algorithm}
\usepackage{algpseudocode}

\usepackage{amssymb}
\usepackage{cite}
\usepackage{stfloats}

\usepackage{amsthm}
\usepackage{float}
\usepackage{amsthm}
\usepackage{amssymb}

\usepackage{graphicx}%
\usepackage{acronym}
\usepackage{balance}
\usepackage{stfloats}
\usepackage{color}
\usepackage{subcaption}

\usepackage{multirow}

\newtheoremstyle{myremark}%
  {}
  {}
  {}
  {\parindent}
  {\itshape}
  {:}
  {5pt plus 1pt minus 1pt}
  {\thmname{#1}\thmnumber{~#2}\thmnote{~(#3)}}

\theoremstyle{myremark}
\newtheorem{theorem}{Theorem}
\newtheorem{lemma}{Lemma}
\theoremstyle{myremark}
\newtheorem*{remark}{Remark}

\acrodef{CRN}{cognitive radio network}
\acrodef{SU}{secondary user}
\acrodef{PU}{primary user}
\acrodef{ZF}{zero forcing}
\acrodef{FD}{full-duplex}
\acrodef{BS}{base station}
\acrodef{i.i.d.}{independent and identically distributed}
\acrodef{DL}{downlink}
\acrodef{UL}{uplink}
\acrodef{SINR}{signal-to-interference-plus-noise ratio}
\acrodef{SNR}{signal noise ratio}
\acrodef{AWGN}{additive white Gaussian noise}
\acrodef{MMSE}{minimum mean square error}
\acrodef{SIC}{successive interference cancellation}
\acrodef{SI}{self-interference}
\acrodef{CCI}{co-channel interference}
\acrodef{MUI}{multiuser interference}
\acrodef{NOMA}{non-orthogonal multiple access}
\acrodef{OMA}{orthogonal multiple access}
\acrodef{QoS}{quality-of-service}
\acrodef{SIC}{successive interference cancellation}
\acrodef{SVD}{singular value decomposition}
\acrodef{MIMO}{multiple-input multiple-output}
\acrodef{SISO}{single-input single-output}
\acrodef{MIMO-NOMA}{multiple-input multiple-output non-orthogonal multiple access}
\acrodef{MIMO-OMA}{multiple-input multiple-output orthogonal multiple access}
\acrodef{PA}{power allocation}

\begin{document}

\title{{\color{black}Energy-Efficient Joint User-RB Association and Power Allocation for Uplink Hybrid NOMA-OMA}}

%
%
%

\author{\IEEEauthorblockN{Ming Zeng, \emph{Student Member, IEEE},
Animesh Yadav, \emph{Member, IEEE}, Octavia A. Dobre, \emph{Senior Member}, \emph{IEEE}, and H. Vincent Poor, \emph{Fellow, IEEE} 
} }

\maketitle

\begin{abstract}
In this paper, energy efficient resource allocation is considered for an uplink hybrid system, where \ac{NOMA} is integrated into \ac{OMA}. To ensure the quality of service for the users, a minimum rate requirement is pre-defined for each user. We formulate an energy efficiency (EE) maximization problem by jointly optimizing the user clustering, channel assignment and power allocation. To address this hard problem, {\color{black}a many-to-one bipartite graph is first constructed considering the users and resource blocks (RBs) as the two sets of nodes. Based on swap matching, a joint user-RB association and power allocation scheme is proposed, which converges within a limited number of iterations. Moreover, for the power allocation under a given user-RB association, we first derive the feasibility condition. If feasible, a low-complexity algorithm is proposed, which obtains optimal EE under any \ac{SIC} order and an arbitrary number of users. In addition, for the special case of two users per cluster, analytical solutions are provided for the two \ac{SIC} orders, respectively. These solutions shed light on how the power is allocated for each user to maximize the EE. Numerical results are presented, which show that the proposed joint user-RB association and power allocation algorithm outperforms other hybrid multiple access based and OMA-based schemes.
}
\end{abstract}

\begin{IEEEkeywords}
Non-orthogonal multiple access (NOMA), energy efficiency (EE), power allocation (PA), uplink transmission.  
\end{IEEEkeywords}

\IEEEpeerreviewmaketitle

\section{Introduction}
{\let\thefootnote\relax\footnote{
Copyright (c) 2015 IEEE. Personal use of this material is permitted. However, permission to use this material for any other purposes must be obtained from the IEEE by sending a request to pubs-permissions@ieee.org.

This work was supported in part by the Natural Sciences and Engineering Research Council of Canada (NSERC) through its Discovery program, and in part by the U.S. National Science Foundation under Grant CCF--1513915. This work has been presented in part at Globecom 2018 \cite{Ming_GC18}.

M. Zeng and O. A. Dobre are with Memorial University, St. John's, NL A1B 3X9, Canada (e-mail: mzeng, odobre@mun.ca).

A. Yadav is with Syracuse University, Syracuse, NY 13210 USA (e-mail: ayadav04@syr.edu).

H. V. Poor is with Princeton University, Princeton, NJ 08544 USA (e-mail: poor@princeton.edu)}  
}
%
Non-orthogonal multiple access (NOMA)\acused{NOMA} has been considered as a promising candidate for the fifth generation (5G) and beyond 5G cellular networks\cite{RA, 16, 23, 2, 3, Zhou_Network18}. The key idea of NOMA is to serve multiple users simultaneously over the same radio resources. The introduced inter-user interference is mitigated by employing \acf{SIC} at the receiver. Downlink \ac{NOMA} has been extensively studied so far. Some works target sum rate maximization and show that higher spectral efficiency (SE) can be achieved by \ac{NOMA} when compared with conventional \acf{OMA}{\color{black}\cite{12, 9, 18, 25, Di_WCOM16, Fair}}. Other works study energy efficiency (EE) maximization and show that \ac{NOMA} can also deliver higher EE than \ac{OMA}\cite{20, 21, 22, EE, Hao_IoT18}. {\color{black}In addition, NOMA has also been applied to downlink celluar machine-to-machine (M2M) communication, and it is shown that improved outage probability can be achieved by NOMA when compared with OMA \cite{Lv_IoT18}.}

While uplink NOMA has been less studied compared with downlink NOMA, {it} has been gaining more attention recently {\color{black}\cite{Endo_ISWCS12, Chen_VTC14, Liu_2018, NORA, RNOMA, joint, Dynamic, Zhai_WCL18}. In \cite{Endo_ISWCS12} and \cite{Chen_VTC14}, system-level throughput performance is studied, and it is shown that compared with OMA, enhanced performance can be obtained by NOMA through proportional fair-based scheduling and fractional transmission power control. \cite{Liu_2018} proposes to incorporate multi-level received power and sequence grouping into existing NOMA schemes, and shows that the proposed scheme can support larger connectivity and higher reliability.} In terms of connectivity, since \ac{SIC} is conducted at the \ac{BS}, which is less complexity- and energy-constrained, uplink NOMA can support more users than the downlink case. This makes it a promising candidate for providing massive connectivity for the {Internet-of-Things} (IoT) \cite{NORA, RNOMA}. \cite{NORA} proposes a non-orthogonal random access (NORA) scheme based on \ac{SIC} to alleviate the access congestion problem facing IoT. {Analytical} and simulation results verify the superiority of NORA over OMA in terms of the preamble collision probability, access success probability, and throughput. \cite{RNOMA} also considers a random NOMA strategy for massive IoT, and derives system stability conditions for the maximum packet arrival rate with and without \ac{QoS} guarantee. However, in the above works on random access, since the focus is on system stability, collision probability, and throughput, quite simple power allocation (PA) algorithms are used, e.g., in \cite{NORA}, the power back-off parameter is simply based on the index of the \ac{SIC} order.

Owing to the vital role of PA in uplink NOMA, such as affecting the rate distribution among {users}, and determining their channel access, it deserves further study. In uplink NOMA, the SIC receiver requires diverse arrived power levels
to distinguish user signals. This is quite different from OMA systems, in which an equal arrived power is desired
by the \ac{BS} to provide uniform \ac{QoS}. 
In \cite{joint}, joint power control and beamforming is studied to maximize the system sum rate for millimeter-wave communications. A sub-optimal solution is proposed, and simulation results show that the proposed solution achieves a close-to-bound uplink sum-rate performance. {\color{black}However, it only applies to single carrier system with two users. {\color{black}The authors in \cite{Dynamic} consider} a multi-carrier system, in which each subcarrier can support multiple users. A greedy user clustering algorithm is first proposed based on users' channel gains. Then, closed-form power allocation solutions are derived. {\color{black}However, \cite{Dynamic} is based on the strong and impractical assumption that each user has the same channel gain over different subcarriers. This is overcome by \cite{Zhai_WCL18}, in which the authors first derive the optimal {PA} under given channel assignment, and then propose a low-complexity joint channel assignment and power allocation using maximum weighted independent set in graph theory.     
Nonetheless, the proposed solution in \cite{Zhai_WCL18} only supports two users on a subcarrier.}}

The aforementioned PA schemes are for SE maximization. With EE becoming a major concern for 5G, studying PA under EE is of importance, especially for power-constrained user equipments \cite{Lv_IoT182}. The energy minimization of NOMA for uplink cellular M2M communications is studied in \cite{EE_NOMA_M2M}, where it is shown that transmitting with minimum rate and full time minimizes the energy consumption. In \cite{Ming_TVT19}, energy-efficient PA for uplink mmWave massive MIMO system with NOMA is studied, and it is shown that NOMA can deliver  superior EE when compared with OMA. Note that \cite{Ming_TVT19} also only allows two users to form a NOMA cluster. Different from previous works, in this paper, {\color{black}we study the EE of an uplink hybrid system with NOMA integrated into OMA (HMA) to support a larger number of IoT devices. The reasons for adopting the HMA system instead of simply applying NOMA among all IoT devices are as follows: 1) the IoT devices may not be able to process over the whole available bandwidth; 2) the delay introduced in decoding the superposed signals may be too large for the IoT-based application; note that for both NOMA and OMA, the total number of decoding is the same, but NOMA has to be done {sequentially}, while OMA can be done in parallel; 3) the error propagation in SIC can become severe {\color{black}as the number of users increases}. The system objective is to} maximize the EE of the considered system under an arbitrary number of users, each with a minimum rate requirement. 

The considered EE maximization problem requires a joint consideration of user clustering, channel assignment and {PA}, which is non-convex and challenging to handle. To tackle it, a many-to-one bipartite graph is first constructed, considering the users and resources blocks (RBs) as the two sets of nodes. Then, based on swap matching, we propose a joint user-RB association and {PA} algorithm, which is guaranteed to converge. Moreover, regarding the power allocation under a given user-RB association, we first derive its feasibility conditions. If feasible, the considered problem is shown to be pseudo-concave and a low-complexity algorithm is proposed, which can obtain optimal EE for any \ac{SIC} order and an arbitrary number of users. Moreover, to further shed light on how the power is allocated for each user to maximize the EE, we derive analytical solutions for the special case of two users per cluster for the two \ac{SIC} orders, respectively, by exploiting the property of pseudo-concave function. Extensive numerical simulations are performed, which validate the superiority of the proposed joint user-RB association and {PA} scheme over other HMA- and OMA-based algorithms.


The rest of the paper is organized as follows: Section II introduces the system model and problem formulation{\color{black}; Section III presents} the proposed  joint user-RB association and power allocation scheme; Section IV shows the proposed low-complexity optimal {PA} algorithm under a given user-RB association; Section V discusses the special case of two users per cluster; Section VI shows the simulation results; {\color{black}Section VII finally draws the conclusions.

\section{System Model and Problem Formulation}
\subsection{System model}
{\color{black}In this paper, uplink is considered, in which a set of users denoted by $\mathcal{U} = \{1,\cdots, U\} $ require to simultaneously access the \ac{BS}}. The overall system bandwidth is $B$ Hz, which is equally divided into $M$ resource blocks (RBs), each with $\frac{B}{M}$ Hz. It is assumed that each RB can accommodate multiple users by employing NOMA, while each user can access only one RB{\color{black}\cite{20, Dynamic, Zhai_WCL18}. The considered scheme has the flavor of both NOMA and OMA techniques, and is thus, referred to as HMA}. {\color{black}Users sharing the same RB form a cluster}. Considering user fairness, the number of users accommodated by the $m$th RB is given by $L_m=\lceil \frac{U}{M} \rceil-1$ or $\lceil \frac{U}{M} \rceil, \forall m \in \{1,\cdots, M\}$, and $\sum_{m=1}^M L_m=U$. Here, we assume that {\color{black}user-RB association} is already performed for the sake of presentation, i.e., user $(m,l), l \in \{1,\cdots,L_m\}$ means the $l$th user in the $m$th RB. The way of conducting user-RB association will be presented in the next section. 
{\color{black}Let us denote} the channel of user $(m,l)$ as $h_{m,l}$, which is characterized by large scale path-loss and small scale Rayleigh fading. Without loss of generality, we also assume that the users' channels are arranged in a descending order on each {\color{black}RB}: $|h_{m,1}| \geq \cdots \geq |h_{m,L_m}|, \forall m \in \{1, \cdots M\}$. According to the NOMA protocol, the received signal at the BS {\color{black}on RB $m$} is given by 
}


{\color{black}
\begin{equation} \label{received signal}
y_m= \sum_{l=1}^{L_m} \sqrt{P_{m,l} }h_{m,l}  s_{m,l} +  n_m,
\end{equation}
where} $s_{m,l}$ denotes the signal transmitted from the $l$th user over the $m$th RB, satisfying ${\mathbb{E}(|s_{m,l}|^2)}=1$. In addition, $P_{m,l}$ denotes the corresponding transmit power, satisfying $P_{m,l}\leq P_{m,l}^{\rm{max}}$, where $P_{m,l}^{\rm{max}}$ is the maximum transmit power for user $(m,l)$. $n_m$ denotes the additive white Gaussian noise (AWGN) at the $m$th RB, which is of zero-mean and variance $\sigma^2$. Different from downlink, all received signals at the \ac{BS} are desired signals in uplink, {\color{black}although there is multiuser interference}.

In downlink, the SIC order is fixed and follows the ascending order of the channel gains, i.e., the users with lower channel gains are decoded first and removed. In contrast, in uplink, the SIC order can be flexible as all received signals at the \ac{BS} are desired signals, e.g., the \ac{BS} can choose to decode the user in an arbitrary order. However, regardless of that, in order to apply SIC and decode signals at the BS, PA should be fully exploited such that the distinctness among various signals is maintained. As a result, conventional PA strategies for OMA (typically intended to equalize the
received signal powers for all users) are not suitable for uplink NOMA systems.
For the sake of analysis, here we assume that the SIC order which decodes user 1 first is employed at the BS. Note that the developed analytical results can be easily extended to other SIC orders. Also, for the special case of two users per cluster, the corresponding two SIC orders are explicitly studied later in the paper. According to the NOMA protocol, the achievable rate (bit/s/Hz) for user $(m,l)$ can be expressed as 
\begin{equation}\label{eq:rate}
R_{m,l}=\log_2 \left(
1+ \frac{P_{m,l} |h_{m,l}|^2 }
{\sum_{k=l+1}^{L_m} P_{m,k} |h_{m,k}|^2+ \sigma^2}
\right),
\end{equation} 
where $\sum_{k=l+1}^{L_m} P_{m,k} |h_{m,k}|^2$ denotes the inter-user interference after SIC. {Particularly, when $k=L_m$, we assume $\sum_{k=L_m+1}^{L_m} P_{m,k} |h_{m,k}|^2=0$, i.e., user $(m,L_m)$ receives no interference from other users.}

\subsection{Problem formulation}
The objective is to maximize the EE of the considered system
while guaranteeing a minimum QoS for each user, i.e., $R_{m,l} \geq R_{m,l}^{\rm{min}}$. Note that in uplink, since users are constrained by their own individual maximum transmit power, and only receive interference from users in the same cluster due to orthogonal resources assigned to each cluster, each user {\color{black}may not concern} the whole system EE, but its own cluster EE. {\color{black}Nonetheless}, with multiple users and RBs, we need to consider the system EE by appropriately pairing the users and assigning the RBs. 

The EE for each cluster is defined as the ratio of the achievable cluster sum rate over the total consumed power \cite{21}. The achievable cluster sum rate is {\color{black}given by $R_m^{\rm{sum}}=\sum_{l=1}^{L_m} R_{m,l}$}, while the total power consumption includes two parts: the fixed circuit power consumption $P_m^{\rm{f}}$ and the flexible transmit power $P_m^{\rm{t}}= \sum_{l=1}^{L_m} P_{m,l}$. Therefore, the EE for the $m$th cluster is given by  
\begin{equation} \label{eq:EE_def}
\eta_m^{\rm{EE}}=\frac{R_m^{\rm{sum}}} { P_m^{\rm{f}} +P_m^{\rm{t}}}.
\end{equation}


Accordingly, the considered problem can be formulated as
\begin{IEEEeqnarray*}{clr}\label{eq:EE}
\displaystyle {\underset{P_{m,l}}{\rm{max}} }  &~ \eta_{\rm{S}}^{\rm{EE}} \IEEEyesnumber \IEEEyessubnumber* \\  
\text{s.t.}  &~ R_{m,l} \geq R_{m,l}^{\text{min}}, {\forall m}, l\in\{1,\cdots, L_m\} \label{eq:b} \\
&~ P_{m,l}  \leq P_{m,l}^{\rm{max}}, {\forall m}, l\in\{1,\cdots, L_m\}, \label{eq:c}
\end{IEEEeqnarray*}
where $\eta_{\rm{S}}^{\rm{EE}}=\sum_{m=1}^M \eta_m^{\rm{EE}}$ denotes the system EE. \eqref{eq:b} and \eqref{eq:c} denote the \ac{QoS} requirement and the transmit power constraint for each user, respectively.

{\color{black} 

{\color{black}\section{Joint User-RB Association and Power Allocation (PA)}}
As the considered system is hybrid, we need to associate the users with the RBs, and allocate the power. However, deriving an optimal joint user-RB association and {PA} scheme is challenging owing to the intra-cluster interference among users. {Indeed, changing the association of a user from one RB to another not only influences this user, but also affects the other users in these RBs}. Moreover, the objective function (\ref{eq:EE}a) is non-convex, which makes it difficult to derive conditions for optimality. 

{\color{black}\subsection{Proposed algorithm}}
{\color{black}To develop a low-complexity joint user-RB association and PA algorithm, we consider the users and RBs as two sets of nodes in a bipartite graph. Then, the objective is to match the users to the RBs and allocate power appropriately such that the EE can be maximized. First, we define a matching as an assignment of RBs to users as follows.

Definition 1: Given two disjoint sets, $\mathcal{U} = \{1,\cdots, U\} $ of
the users, and $\mathcal{M} = \{1,\cdots, M\} $ of the RBs, a many-to-one matching $\Phi$ is a mapping from the set $\mathcal{U} \cup \mathcal{M}$ into the set of all subsets of $\mathcal{U} \cup \mathcal{M}$ such that for every $u \in \mathcal{U}$ and $m \in \mathcal{M}$:
\begin{enumerate}
\item $\Phi(u) \in \mathcal{M}$;
\item $\Phi(m) \subseteq \mathcal{U}$;
\item $|\Phi(u)|=1$;
\item $|\Phi(m)|=L_m$;
\item $m = \Phi(u) \Leftrightarrow u \in \Phi(m)$,
\end{enumerate} 
{\color{black}where $|\cdot|$ returns the size of the matching.} Conditions $1)$ and $3)$ state that each user is matched with an RB, while conditions $2)$ and $4)$ imply that each RB is matched with $L_m$ users. 

Inspired by the many-to-one housing assignment problem \cite{bodine2011peer}, we introduce the notion of swap matching into our many-to-one matching model, and propose a matching algorithm for the joint user-RB association and PA problem{\color{black}\cite{Di_WCOM16}}. A swapping operation means two users matched with different RBs exchange their matches, while the matching for other users remains the same.
{\color{black}The PA is then updated for the two corresponding RBs. Note that how to allocate power to obtain the EE for a given RB will be presented in the following sections, and we assume it is known here.
To ensure an improved EE performance,
a swapping operation is approved and the matching is updated only when the sum of the EEs for the two RBs involved increases after the swap. Then, to maximize the EE of the considered system,} the idea is to continue the swapping operation until no swapping is further approved. Pseudocode for the proposed swapping-based algorithm is given in Algorithm 1.  

Note that in the initialization phase, the basic idea is to associate the user to the RB in which it has a large channel gain. This leads to either a higher data rate for the user, or a lower transmit power. Both yield a higher EE. Then, for the swap matching phase, iterations will continue until no swapping operation can be approved in a new round. 

\subsection{Convergence and complexity}
\begin{theorem}
The proposed joint user-RB association and PA algorithm converges after a finite number of swapping operations.
\end{theorem}

\begin{IEEEproof}
After a number of swapping operations, the structure of matching changes as follows:
\begin{equation}
\Phi_0 \rightarrow \Phi_1 \rightarrow \Phi_2 \rightarrow \cdots,
\end{equation}
{\color{black}where $\Phi_0$ is the initial matching}. For swapping operation $l$, the matching changes from $\Phi_{l-1}$ to $\Phi_{l}$. Denote the corresponding system EE as $\eta_{\rm{S}}^{\rm{EE}} (l-1)$ and $\eta_{\rm{S}}^{\rm{EE}} (l)$. Therefore, we have $\eta_{\rm{S}}^{\rm{EE}} (l) > \eta_{\rm{S}}^{\rm{EE}} (l-1)$, i.e., the system EE increases at each swapping operation. Moreover, the system EE clearly has an upper bound due to the limited power and spectrum resources. Consequently, the number of potential swapping operations is finite. 
\end{IEEEproof}

{\color{black}
\begin{algorithm}
\caption{{\color{black}Proposed joint user-RB association and PA algorithm.}}\label{BM}
{\color{black}
\begin{algorithmic}[1]
\State {\textbf{Step 1: \textit{Initialization phase}}}
\State $K \leftarrow \lceil \frac{U}{M} \rceil$, $\hat{U}\leftarrow U$;
\State \textbf{for} $k = \{1,\cdots,K\}$
\State \hspace{10pt} $\hat{M}\leftarrow M$, $\text{Count} \leftarrow 1$;
\State \hspace{10pt} \textbf{while} ($\text{Count} \leq M$)
\State \hspace{10pt} \hspace{10pt} $h_{m^{\star},u^{\star}} \leftarrow \max \{ |h_{m,u}| \}|_{\forall m \in \hat{M}, \forall u \in \hat{U}}$
\State \hspace{10pt} \hspace{10pt} assign $u^{\star}$ to RB $m^{\star}$;
\State \hspace{10pt} \hspace{10pt} $\hat{U}\leftarrow \hat{U} \backslash u^{\star}$, $\hat{M}\leftarrow \hat{M} \backslash m^{\star}$; 
\State \hspace{10pt} \hspace{10pt} $\text{Count}  \leftarrow \text{Count}+1$;
\State \hspace{10pt} \textbf{end while}
\State \textbf{end for}
\State {\textbf{Step 2: \textit{Swap matching phase}}}
\State $Indicator=1$;
\State \textbf{while} $(Indicator)$
\State \hspace{10pt} $Indicator=0$;
\State \hspace{10pt} \textbf{for} $u \in \{1,\cdots, U\} $,
\State \hspace{10pt} \hspace{10pt} \textbf{for} $k \in \{1,\cdots, U\}$ 
\State \hspace{10pt} \hspace{10pt} \hspace{10pt} \textbf{if} $\Phi(k)=\Phi(u)$
\State \hspace{10pt} \hspace{10pt} \hspace{10pt} \hspace{10pt} continue;
\State \hspace{10pt} \hspace{10pt} \hspace{10pt} \textbf{else} 
\State \hspace{10pt} \hspace{10pt} \hspace{10pt} \hspace{10pt} calculate and compare the EE before and after the swap using Algorithm 2; 
\State \hspace{10pt} \hspace{10pt} \hspace{10pt} \hspace{10pt}\textbf{if} EE increases
\State \hspace{10pt} \hspace{10pt} \hspace{10pt} \hspace{10pt} \hspace{10pt} update the matching, $Indicator \leftarrow 1$;
\State \hspace{10pt} \hspace{10pt} \hspace{10pt} \hspace{10pt} \textbf{end if}
\State \hspace{10pt} \hspace{10pt} \hspace{10pt} \textbf{end if}
\State \hspace{10pt} \hspace{10pt} \textbf{end for}
\State \hspace{10pt} \textbf{end for}
\State \textbf{end while}
\end{algorithmic}
}
\end{algorithm}
}

Given the convergence of the proposed algorithm, we discuss its computational complexity. For the initial phase, it takes $O(U^2M)$ operations. In the swap matching phase, denote the number of iterations to reach the final matching as $I_1$.\footnote{{\color{black}This number cannot be given in closed form, since we do not know for sure at which iteration the proposed algorithm reaches the final matching. This is quite common in the design of most heuristic algorithms. To evaluate the convergence speed, we will show the distribution of this number in the Simulation Results section.}} In each iteration, all possible swapping combinations should be considered, which requires $O(U^2)$ operations. In each swapping attempt, we need to conduct PA to calculate the EE before and after the swapping for the two related clusters. Denote the computational complexity of the power allocation for calculating the EE as $O(X)$, which will be given in the following section. Then, the total complexity for the swap matching phase is $O(I_1 U^2X)$. Adding this to the initial phase, we obtain the total complexity as $O(U^2(I_1 X+M))$.

}



}

{\color{black}
\section{Power Allocation under Given User-RB Association}
In line 21 of Algorithm 1, we assume that the way of allocating power under a given user-RB association is known. In this section, we present in detail how we conduct PA to maximize the EE. 
Under a given user-RB association, we can conclude that maximizing the system EE is equivalent to maximizing the EE for each cluster. This is because the system EE is the summation over all cluster EEs, which are mutually independent as they are allocated with different RBs.
As a result, we can consider the EE maximization problem on each RB separately, and the $m$th subproblem is given by }
\begin{IEEEeqnarray*}{clr}\label{eq:EE_cluster}
\displaystyle {\underset{P_{m,l}}{\rm{max}} }  &~ \eta_m^{\rm{EE}} \IEEEyesnumber \IEEEyessubnumber* \\  
\text{s.t.}  &~ R_{m,l} \geq R_{m,l}^{\text{min}}, l\in\{1,\cdots, L_m\} \label{cluster:b} \\
&~ P_{m,l}  \leq P_{m,l}^{\rm{max}},  l\in\{1,\cdots, L_m\}. \label{cluster:c}
\end{IEEEeqnarray*}

{{As the considered subproblems have the same form on different RBs, in the following sections, we omit the RB index $m$ for notational simplicity. Also, $L_m$ is replaced by $L$ while $\eta_m^{\rm{EE}}$ is replaced by $\eta_{\rm{EE}}$.}} 

\subsection{Determine the feasibility}
Owing to the minimum rate requirements and transmit power constraints, \eqref{eq:EE_cluster} may be infeasible, i.e., there may not exist a PA solution to satisfy all the constraints. As a result, we need to find the feasibility conditions first. Observe that the last user receives no interference from other users due to SIC; we start with it and obtain 
\begin{align} \label{rate:M}
& \log_2 \left(
1+ \frac{P_{L} |h_{L}|^2 }
{\sigma^2}
\right) \geq R_{L}^{\text{min}} \\ \nonumber
 \Leftrightarrow & P_{L} \geq \frac{(2^{R_{L}^{\text{min}}}-1)}{|h_{L}|^2}.
\end{align}

To satisfy the above requirement, we have
\begin{equation} \label{L}
P_{L}^{\rm{max}} \geq \frac{(2^{R_{L}^{\text{min}}}-1)}{|h_{L}|^2}.
\end{equation}

Assume that \eqref{L} is satisfied. Clearly, to reduce the interference from user $L$ to other users, {\color{black}it needs to use the minimum transmit power}, i.e., $P_{L}=P_{L}^{\rm{min}} = \frac{(2^{R_{L}^{\text{min}}}-1)}{|h_{L}|^2}$. Now we consider the $(L-1)$th user. Likewise, we have
\begin{align} \label{rate:M-1}
&\log_2 \left(
1+ \frac{P_{L-1} |h_{L-1}|^2 }
{P_{L} |h_{L}|^2+\sigma^2}
\right) \geq R_{L-1}^{\text{min}}  \nonumber \\
 &\Leftrightarrow P_{L-1} \geq \frac{2^{R_{L}^{\text{min}}}(2^{R_{L-1}^{\text{min}}}-1)}{|h_{L-1}|^2} \\
 & \Rightarrow P_{L-1}^{\rm{max}}  \geq \frac{2^{R_{L}^{\text{min}}}(2^{R_{L-1}^{\text{min}}}-1)}{|h_{L-1}|^2}.  \nonumber
\end{align}

Using the mathematical induction, we can easily extend it to all users, and obtain
\begin{equation}
P_{l}^{\rm{max}}  \geq P_{l}^{\rm{min}}= \frac{2^{\sum_{k=l+1}^L R_{k}^{\text{min}}}(2^{R_{l}^{\text{min}}}-1)}{|h_{l}|^2}, \forall l \in \{1, \cdots, L\},
\end{equation}
where $P_{l}^{\rm{min}}$ is the minimum power required to satisfy the minimum rate requirement for the $l$th user. {Here we assume that $\sum_{k=L+1}^L R_{k}^{\text{min}}=0$.} Once the above conditions between the minimum rate requirements and the power constraints are satisfied, \eqref{eq:EE_cluster} is feasible. 

\subsection{Maximizing the EE when \eqref{eq:EE_cluster} is feasible}
The objective function (\ref{eq:EE_cluster}a) is of fractional form, which is non-convex and challenging to handle. To tackle it, we first deal with the numerator, i.e., the sum rate, which can be re-written as {\color{black}
\begin{align} \label{sum rate}
R^{\rm{sum}}& =\sum_{l=1}^{L} \log_2 \left(
1+ \frac{P_{l} |h_{l}|^2 }
{\sum_{k=l+1}^{L} P_{k} |h_{k}|^2+ \sigma^2} \right)  \nonumber \\ 
&= \log_2 \left( 
1+ \frac{\sum_{l=1}^{L} P_{l} |h_{l}|^2 }
{\sigma^2}
\right).
\end{align}
}

It is easy to see that the sum rate is a concave function with
respect to (w.r.t.) the transmit power for each user. 

Now we consider the QoS constraints (\ref{eq:EE_cluster}b). It is non-
convex on its current form. However, after some mathematical
manipulations, it can be reformulated as
\begin{equation} \label{minimum rate}
P_{l} |h_{l}|^2 \geq \left( 2^{R_{l}^{\text{min}}}-1 \right) \left(\sum_{k=l+1}^{L} P_{k} |h_{k}|^2+\sigma^2 \right),
\end{equation}
which is a linear constraint, since it is just an affine mapping w.r.t., $P_{l}, l\in \{1, \cdots, L\}$. 

Accordingly, problem \eqref{eq:EE_cluster} can be re-written as
\begin{IEEEeqnarray*}{clr}\label{eq:EE_new}
\displaystyle {\underset{P_{l}}{\rm{max}} }  &~ \frac{\log_2 \left( 
1+ \frac{\sum_{l=1}^{L} P_{l} |h_{l}|^2 }
{\sigma^2}
\right)}{P_f + \sum_{l=1}^{L} P_{l}} \IEEEyesnumber \IEEEyessubnumber* \\  
\text{s.t.}  &~ \eqref{minimum rate}~ \eqref{cluster:c}, l\in\{1,\cdots, L\} .
\end{IEEEeqnarray*}

For the objective function (\ref{eq:EE_new}a), its numerator is a strictly concave function w.r.t., $P_{l}, ~l\in \{1, \cdots, L\}$, while the denominator is an affine mapping over $P_{l}, ~l\in \{1, \cdots, L\}$. Therefore, it is a strictly pseudo-concave function \cite[Proposition 6]{24}. According to the property of strictly pseudo-concave function, it can be optimally solved by applying the Dinkelbach's algorithm \cite[Proposition 6]{24}. The specific procedure is summarized in Algorithm \ref{Dinkelbach}. Denote the number of iterations for Algorithm 2 to converge as $I_2$. During each iteration, the proposed algorithm needs to solve {\color{black}the following problem, i.e., line 4,
\begin{IEEEeqnarray*}{clr}\label{eq:SE_new}
\displaystyle {\underset{P_{l}}{\rm{max}} }  &~ {\log_2 \left( 
1+ \frac{\sum_{l=1}^{L} P_{l} |h_{l}|^2 }
{\sigma^2} \right)}-\beta (P_f + \sum_{l=1}^{L} P_{l}) \IEEEyesnumber \IEEEyessubnumber* \\  
\text{s.t.}  &~ \eqref{minimum rate}~ \eqref{cluster:c}, l\in\{1,\cdots, L\},
\end{IEEEeqnarray*}
where $\beta$ is known. Clearly, the above problem is concave, and can be solved using standard {\color{black}algorithms}, such as interior-point method. However, the standard {\color{black}approach} does not exploit the specific structure of \eqref{eq:SE_new}, and is computationally intensive when \eqref{eq:SE_new} needs to be solved over and over again. This is indeed the case here, since 
solving \text{Algorithm 1} requires solving Algorithm 2 many times, i.e., line 21, and addressing Algorithm 2 also requires to solve \eqref{eq:SE_new} many times, i.e., \text{line 4}. To relieve the computational burden, we propose a low-complexity optimal solution for \eqref{eq:SE_new} as follows: 


Denote $F={\log_2 \left( 
1+ \frac{\sum_{l=1}^{L} P_{l} |h_{l}|^2 }
{\sigma^2} \right)}-\beta (P_f + \sum_{l=1}^{L} P_{l})$. Then, for user $l$, we have $\frac{\partial F}{\partial P_l}= \frac{|h_{l}|^2}{\ln2 (\sum_{k=1}^{L} P_{k} |h_{k}|^2 + \sigma^2)} - \beta$. Setting $\frac{\partial F}{\partial P_l}=0$, we obtain $P_l^{0}=\frac{1}{\beta \ln2}- \frac{\sum_{k \neq l}P_k |h_{k}|^2 + \sigma^2}{|h_{l}|^2}$. 
If all other power values, i.e., $P_k, k\neq l$ are fixed, we can easily obtain the optimal solution for user $l$ by comparing $P_l^{0}$ with its minimum and maximum power constraints. Specifically, the optimal power $P_l^{\star}$ is given by
\begin{equation} \label{power_update}
P_l^{\star}=
\begin{cases}
P_{l}^{\rm{min}}, & \text{if $P_l^{0}<P_{l}^{\rm{min}}$},\\
P_{l}^{\rm{max}},& \text{if $P_l^{0}>P_{l}^{\rm{max}}$}, \\ 
P_l^{0}, & \text{otherwise}.
\end{cases}
\end{equation}

On this basis, the proposed low-complexity algorithm goes as follows: we first allocate the minimum required power to each user; then, we update the power for the users one by one using \eqref{power_update}; this update continues until convergence. Note that convergence is guaranteed since $F$ increases or remains unchanged after each update, and there exists an upper bound. Denote the number of iterations for convergence as $I_3$; then, its complexity is just $O(I_3)$. Thus, we have {\color{black}$X=I_2 I_3 $}, and $O(U^2(I_1 X+M))=O(U^2(I_1 I_2 I_3+M))$. Moreover, the obtained local optimum is also the global optimum since \eqref{eq:SE_new} is concave. The specific procedure is summarized in \text{Algorithm 3}.


}


{\color{black}
\begin{remark}
For any other SIC order, it can be easily proved that the objective function is the same as (\ref{eq:EE_new}a). Moreover, the minimum rate constraints can be turned into convex constraints similar to \eqref{minimum rate}. Therefore, Algorithms 2 and 3 can be directly used for EE maximization under any other SIC order.  
\end{remark}
}

\begin{algorithm} 
\caption{Proposed EE maximization PA algorithm.} \label{Dinkelbach}
\begin{algorithmic}[1]
\State {\textbf{Initialize parameters.}}
\State \hspace{10pt} Set $\epsilon>0; \beta \leftarrow 0; F> \epsilon;$ 

\State {\textbf{while} $F>\epsilon$ \textbf{do}}
	\State \hspace{10pt} $P_{l}^{\star} \leftarrow  \text{argmax}~ {\log_2 \left( 
1+ \frac{\sum_{l=1}^{L} P_{l} |h_{l}|^2 }
{\sigma^2} \right)}-\beta (P_f + \sum_{l=1}^{L} P_{l})$; s.t. \eqref{minimum rate}~ \eqref{cluster:c};
	\State \hspace{10pt} $F \leftarrow {\log_2 \left( 
1+ \frac{\sum_{l=1}^{L} P_{l}^{\star} |h_{l}|^2 }
{\sigma^2} \right)}-\beta (P_f + \sum_{l=1}^{L} P_{l}^{\star})$;
	\State \hspace{10pt} $\beta \leftarrow \frac{\log_2 \left( 
1+ \frac{\sum_{l=1}^{L} P_{l}^{\star} |h_{l}|^2 }
{\sigma^2}
\right)}{P_f + \sum_{l=1}^{L} P_{l}^{\star}}$;
\State {\textbf{end while}}
\end{algorithmic}
\end{algorithm}

{\color{black}
\begin{algorithm} 
\caption{{\color{black}Proposed low-complexity algorithm for \eqref{eq:SE_new}.}} \label{low-complexity}
{\color{black}
\begin{algorithmic}[1]
\State {\textbf{Initialize parameters.}}
\State \hspace{10pt} Set $P_l \leftarrow  P_{l}^{\rm{min}}, l \in \{1,\cdots,L\}$;
\State {\textbf{while} $1$ \textbf{do}}
	\State \hspace{10pt} $P_{\rm{old}} \leftarrow P $;
	\State \hspace{10pt} \textbf{for} $l=\{1,\cdots,L\}$;
	\State \hspace{10pt} \hspace{10pt} $P_l^{0}=\frac{1}{\beta \ln2}- \frac{\sum_{k \neq i}P_k |h_{k}|^2 + \sigma^2}{h_{i}|^2}$;
	\State \hspace{10pt} \hspace{10pt} \textbf{if} $P_l^{0}<P_{l}^{\rm{min}}$
	\State \hspace{10pt} \hspace{10pt} \hspace{10pt} $P_l \leftarrow P_{l}^{\rm{min}}$;
	\State \hspace{10pt} \hspace{10pt} \textbf{elseif} $P_l^{0}>P_{l}^{\rm{max}}$
	\State \hspace{10pt} \hspace{10pt} \hspace{10pt} $P_l \leftarrow P_{l}^{\rm{max}}$;
	\State \hspace{10pt} \hspace{10pt} \textbf{else} 
	\State \hspace{10pt} \hspace{10pt} \hspace{10pt} $ P_l \leftarrow P_l^{0}$;
	\State \hspace{10pt} \hspace{10pt} \textbf{end if} 
	\State \hspace{10pt} \hspace{10pt} \textbf{if} $|P_{\rm{old}}-P|<10^{-9}$
	\State \hspace{10pt} \hspace{10pt} \hspace{10pt} break;
	\State \hspace{10pt} \hspace{10pt} \textbf{end if}
	\State \hspace{10pt} \textbf{end for}
\State {\textbf{end while}}
\end{algorithmic}
}
\end{algorithm}
}
Although the proposed Algorithms \ref{Dinkelbach} and \ref{low-complexity} can be used to solve the considered EE maximization problem, {they do} not shed much light into the behaviour of the system, since an iterated algorithm is used. For example, how much power will be employed by the user with the highest channel gain? To this end, several important properties are observed and listed in the sequel:

\begin{lemma}
Transferring power\footnote{Note that here transferring power means one user lowers his transmit power, while another user increases the same amount of transmit power.} from a user with lower channel gain to a user with higher channel gain leads to increased EE.
\end{lemma}
\begin{IEEEproof}
According to (\ref{eq:EE_new}a), when the power transfer happens, the numerator increases owing to the channel gain ordering. On the other hand, the sum transmit power remains unchanged, and thus, the denominator remains unchanged. Therefore, the EE increases as well.
\end{IEEEproof}

\begin{theorem}
If $\frac{\partial \eta_{\rm{EE}}}{\partial P_{1}}|_{P_{1}^{\rm{max}}, \bar{P}_{-1}} \geq 0$, user 1 should transmit at full power to maximize the EE, where $\bar{P}_{-1}= \bar{P}_{2}, \cdots, \bar{P}_{L}$ denotes a feasible PA solution for the other users. 
\end{theorem}

\begin{IEEEproof}
First, when $P_{1}=P_{1}^{\rm{max}}$, the feasible region for the other users is maximized, since the interference from user 1 is cancelled by SIC, and the minimum rate requirement of user 1 is most likely to be satisfied. This means that if there exists a feasible region, $P_{1}=P_{1}^{\rm{max}}$ is inside it. Furthermore, since $\frac{\partial \eta_{\rm{EE}}}{\partial P_{1}}|_{P_{1}^{\rm{max}},~ \bar{P}_{-1}} \geq 0$, then $P_{1}=P_{1}^{\rm{max}}$ maximizes the EE, when $\bar{P}_{-1}$ remains fixed \cite[Proposition 5]{24}. Next, we consider the case when power transfer happens between user 1 and other users. According to Lemma 1, transferring power from user 1 to other users leads to a lower EE. Therefore, this will not happen. This completes the proof. 
\end{IEEEproof}

\begin{table*} [!h]
\caption{PA Solution for Two Users under Case I.} 
\renewcommand{\arraystretch}{1}
\label{Table 1} 
\centering
 \begin{tabular}{l|l|l} 
 \hline  
\bfseries Phases & \bfseries Conditions & \bfseries Solutions \\ [0.5ex] 
 \hline\hline
 Phase I &$\frac{\partial \eta_{\rm{EE}}}{\partial P_1}|_{P_1^{\rm{max}},P_2^{\rm{max}}} \geq \frac{\partial \eta_{\rm{EE}}}{\partial P_2}|_{P_1^{\rm{max}},P_2^{\rm{max}}} \geq 0$ &$P_1 \leftarrow P_1^{\rm{max}}$,  $P_2 \leftarrow \min \left(P_2^{\rm{max}}, \frac{P_1^{\rm{max}}|h_1|^2}{(2^{R_1^{\rm{min}}}-1)|h_2|^2 }- \frac{\sigma^2}{|h_2|^2} \right)$ \\
 \hline
\multirow{2}{4em} {Phase II} & $\frac{\partial \eta_{\rm{EE}}}{\partial P_1}|_{P_1^{\rm{max}},P_2^{\rm{max}}} \geq 0$, $\frac{\partial \eta_{\rm{EE}}}{\partial P_2}|_{P_1^{\rm{max}},P_2^{\rm{max}}} \leq 0$ & $P_1 \leftarrow P_1^{\rm{max}}$, set $P_2^{\star} \leftarrow \frac{\partial \eta_{\rm{EE}}}{\partial P_2}|_{P_1^{\rm{max}}}=0$; if $P_2^{\star} \leq P_2^{\rm{min}} $, then $P_2 \leftarrow  P_2^{\rm{min}}$ \\ 
& $\frac{\partial \eta_{\rm{EE}}}{\partial P_1}|_{P_1^{\rm{max}},P_2^{\rm{min}}} \geq 0$ $\left(P_2^{\rm{min}}= \frac{(2^{R_2^{\rm{min}}}-1)\sigma^2}{|h_2|^2} \right)$ & else $P_2 \leftarrow \min \left( \frac{\partial \eta_{\rm{EE}}}{\partial P_2}|_{P_1^{\rm{max}}}=0, \frac{P_1^{\rm{max}}|h_1|^2}{(2^{R_1^{\rm{min}}}-1)|h_2|^2 }- \frac{\sigma^2}{|h_2|^2} \right)$; \\
 \hline
\multirow{2}{4em} {Phase III}  & $\frac{\partial \eta_{\rm{EE}}}{\partial P_1}|_{P_1^{\rm{max}},P_2^{\rm{max}}} \leq 0$, $\frac{\partial \eta_{\rm{EE}}}{\partial P_2}|_{P_1^{\rm{max}},P_2^{\rm{max}}} \leq 0$,   &  Same as Phase II\\
 & $\frac{\partial \eta_{\rm{EE}}}{\partial P_1}|_{P_1^{\rm{max}},P_2^{\rm{min}}} \geq 0$ $\left(P_2^{\rm{min}}= \frac{(2^{R_2^{\rm{min}}}-1)\sigma^2}{|h_2|^2} \right)$ & \\
 \hline
Phase IV & $\frac{\partial \eta_{\rm{EE}}}{\partial P_1}|_{P_1^{\rm{max}},P_2^{\rm{min}}} \leq 0$, $\frac{\partial \eta_{\rm{EE}}}{\partial P_2}|_{P_1^{\rm{max}},P_2^{\rm{min}}} \leq 0$   &$P_1 \leftarrow \max \left( 
\frac{\partial \eta_{\rm{EE}}}{\partial P_1}|_{P_2^{\rm{min}}}=0, \frac {(2^{R_1^{\rm{min}}}-1)2^{R_2^{\rm{min}}} \sigma^2 }{|h_1|^2} \right)$, $P_2 \leftarrow  P_2^{\rm{min}}$\\
 \hline
\end{tabular}
\end{table*}

\begin{table*} [!h]
\caption{PA Solution for Two Users under Case II.} 
\renewcommand{\arraystretch}{1}
\label{Table 2} 
\centering
 \begin{tabular}{l|l|l} 
 \hline  
\bfseries Phases & \bfseries Conditions & \bfseries Solutions \\ [0.5ex] 
 \hline\hline
 Phase I &$\frac{\partial \eta_{\rm{EE}}}{\partial P_1}|_{P_1^{\rm{max}},P_2^{\rm{max}}} \geq \frac{\partial \eta_{\rm{EE}}}{\partial P_2}|_{P_1^{\rm{max}},P_2^{\rm{max}}} \geq 0$ &$P_2 \leftarrow P_2^{\rm{max}}$,  $P_1 \leftarrow \min \left(P_1^{\rm{max}}, \frac{P_2^{\rm{max}}|h_2|^2}{(2^{R_2^{\rm{min}}}-1)|h_1|^2 }- \frac{\sigma^2}{|h_1|^2} \right)$ \\
 \hline
\multirow{2}{4em} {Phase II}  & $\frac{\partial \eta_{\rm{EE}}}{\partial P_1}|_{P_1^{\rm{max}},P_2^{\rm{max}}} \geq 0$, $\frac{\partial \eta_{\rm{EE}}}{\partial P_2}|_{P_1^{\rm{max}},P_2^{\rm{max}}} \leq 0$   & if $\overline{P}_1^{\star} \leq P_1^{\rm{min}}$, then $P_1 \leftarrow \overline{P}_1^{\rm{min}}$, $P_2 \leftarrow (\overline{P}_1^{\rm{min}}-b)/k$ \\
& & if $\overline{P}_1^{\star} \in [ \overline{P}_1^{\rm{min}}, P_1^{\rm{max}} ]$, then $P_1 \leftarrow \overline{P}_1^{\star}$, $P_2 \leftarrow \overline{P}_2^{\star}$ \\
&  & if $\overline{P}_1^{\star} \geq P_1^{\rm{max}}$, then $P_1 \leftarrow P_1^{\rm{max}}$, $P_2 \leftarrow \frac{\partial \eta_{\rm{EE}}}{\partial P_2}|_{P_1^{\rm{max}}}=0$ \\
 \hline
Phase III & $\frac{\partial \eta_{\rm{EE}}}{\partial P_1}|_{P_1^{\rm{max}},P_2^{\rm{max}}} \leq 0$, $\frac{\partial \eta_{\rm{EE}}}{\partial P_2}|_{P_1^{\rm{max}},P_2^{\rm{max}}} \leq 0$   &same as Phase II\\
 \hline
\end{tabular}
\end{table*}

\begin{theorem}
If $\frac{\partial \eta_{\rm{EE}}}{\partial P_{1}}|_{P_{1}^{\rm{max}}, P_{-1}^{\rm{min}}} \leq 0$, then $P_{l} =P_{l}^{\rm{min}}, l \neq 1$ and $P_{1}= \max \left( 
\frac{\partial \eta_{\rm{EE}}}{\partial P_{1}}|_{P_{-1}^{\rm{min}}}=0, P_{1}^{\rm{min}} \right)$, where $P_{-1}^{\rm{min}} = P_{2}^{\rm{min}}, \cdots, P_{L}^{\rm{min}}$ denotes the minimum required power. 
\end{theorem}

\begin{IEEEproof}
The derivative $\frac{\partial \eta_{\rm{EE}}}{\partial P_{l}}$ is given by
\begin{equation} \label{derivative}
\begin{split}
\frac{\partial \eta_{\rm{EE}}}{\partial P_{l}} =&
\frac{|h_{l}|^2}{(\sigma^2+\sum_{l=1}^{L} P_{l} |h_{l}|^2)(P_f + \sum_{l=1}^{L} P_{l}) \ln2}  \\
&- \frac{\log_2 \left( 
1+ \frac{\sum_{l=1}^{L} P_{l} |h_{l}|^2 }
{\sigma^2}
\right)}{(P_f + \sum_{l=1}^{L} P_{l})^2}.
\end{split}
\end{equation}

Clearly, the derivative $\frac{\partial \eta_{\rm{EE}}}{\partial P_{l}}$ is arranged following the same order of $|h_{l}|^2$, i.e., $\frac{\partial \eta_{\rm{EE}}}{\partial P_{1}} \geq \cdots \geq \frac{\partial \eta_{\rm{EE}}}{\partial P_{l}} \cdots \geq \frac{\partial \eta_{\rm{EE}}}{\partial P_{L}}$. Since $\frac{\partial \eta_{\rm{EE}}}{\partial P_{1}}|_{P_{1}^{\rm{max}}, P_{-1}^{\rm{min}}} \leq 0$, then $\frac{\partial \eta_{\rm{EE}}}{\partial P_{l}}|_{P_{1}^{\rm{max}}, P_{-1}^{\rm{min}}} \leq 0, \forall l \in \{2,\cdots,L\}$. Therefore, all users should reduce their transmit power to increase the EE \cite[Proposition 5]{24}. On the other hand, for all users except user 1, they can only reduce their power to the minimum required power. So, we have $P_{l} =P_{l}^{\rm{min}}, l \neq 1$. Once all the other users' powers are fixed, the EE is maximized at the unique root of $\frac{\partial \eta_{\rm{EE}}}{\partial P_{1}}|_{P_{-1}^{\rm{min}}}=0$ or the boundary point, i.e., $P_{1}^{\rm{min}}$. Combining this, we can conclude that $P_{1}= \max \left( 
\frac{\partial \eta_{\rm{EE}}}{\partial P_{1}}|_{P_{-1}^{\rm{min}}}=0, P_{1}^{\rm{min}} \right)$.
\end{IEEEproof}

\begin{remark}
Note that the condition for Theorem 2 holds when $P_{l}^{\rm{max}}$ are quite small, while that for Theorem 3 holds when $P_{l}^{\rm{max}}$ are quite large. Thus, some good insights for these two extreme cases have been derived. However, for the cases between these two extremes, it is quite complicated to derive analytical results due to the coupling between the \ac{QoS} requirements and power constraints.
\end{remark}

\section{Two User Case}

Although it is challenging to derive the analytical solution for the general case of multiple users, for the special case of two users, this is possible. 
Since the derivatives $\frac{\partial \eta_{\rm{EE}}}{\partial P_l}$ are arranged as $\frac{\partial \eta_{\rm{EE}}}{\partial P_1} \geq \cdots \geq \frac{\partial \eta_{\rm{EE}}}{\partial P_l} \cdots \geq \frac{\partial \eta_{\rm{EE}}}{\partial P_L}$, for the two user case, there are only three cases to consider: $\frac{\partial \eta_{\rm{EE}}}{\partial P_1} \geq \frac{\partial \eta_{\rm{EE}}}{\partial P_2} \geq 0$, $\frac{\partial \eta_{\rm{EE}}}{\partial P_1} \geq 0 \geq \frac{\partial \eta_{\rm{EE}}}{\partial P_2} $ and $0 \geq \frac{\partial \eta_{\rm{EE}}}{\partial P_1} \geq \frac{\partial \eta_{\rm{EE}}}{\partial P_2}$. On the other hand, under different SIC orders, the feasibility region is different, and thus, different PA solutions are required. The two SIC orders need to be discussed separately. In the following, we first consider the case for the SIC order which decodes \text{user 1} first, {\color{black}and refer to it} as Case I. Then, the other case is considered, which is referred to as Case II.

\subsection{Analytical solution when user 1 is decoded first}
In this case, the PA solutions for the two users are listed in Table I.

\begin{IEEEproof}
Refer to Appendix I.
\end{IEEEproof}



\begin{remark}
The bisection method can be used to find the root for the equation $\frac{\partial \eta_{\rm{EE}}}{\partial P_l}=0$, with the complexity of $\log_2(P_l^{\rm{max}}/\delta)$, where $\delta$ denotes the required precision. This is also the dominant computation of obtaining the solution for the EE. According to Table I, when the system is in Phases I, II or III, user 1 always transmits at full power. In Phase IV, user 2 transmits at minimum power. Moreover, from Phase I to Phase IV, {\color{black}we can see how the users react when the maximum allowable transmit power increases. In Phase I, the maximum allowable transmit power is too small, and all the transmit power should be consumed not to violate the \ac{QoS} constraint. In Phases II and III, user 2 should only transmit with the power which ensures both \ac{QoS} and maximum EE. }
\end{remark}

\subsection{Analytical solution when user 2 is decoded first}
In this case, the problem can be formulated as
\begin{IEEEeqnarray*}{clr}\label{eq:EE_new_2}
\displaystyle {\underset{P_1, P_2}{\rm{max}} }  &~ \frac{\log_2 \left( 
1+ \frac{ P_1 |h_{1}|^2 + P_2 |h_{2}|^2}
{\sigma^2}
\right)}{P_f + P_1+P_2} \IEEEyesnumber \IEEEyessubnumber* \\  
\text{s.t.}  &~ P_l \leq P_l^{\rm{max}}, m\in\{1, 2\}, \\
 &~\log_2 \left(
1+ \frac{P_2 |h_{2}|^2 }
{ P_1 |h_{1}|^2+ \sigma^2}\right) \geq R_2^{\rm{min}}, \\
&~  \log_2 \left(
1+ \frac{P_1 |h_{1}|^2 }
{ \sigma^2}\right) \geq R_1^{\rm{min}},
\end{IEEEeqnarray*}
where (\ref{eq:EE_new_2}c) and (\ref{eq:EE_new_2}d) represent the QoS requirements for user 2 and user 1, respectively. Note that (\ref{eq:EE_new_2}a) is the same as (\ref{eq:EE_new}a) for two users. Indeed, for uplink NOMA, if there exists no QoS constraints, the achievable sum rate under any SIC order is the same, and so is the EE. However, with the QoS constraints, the feasibility region of the power may vary under different SIC orders, and thus, leading to different sum rates and EEs. 

The corresponding solution for the above problem is listed in Table II. Note that in this table, we have $\overline{P}_1^{\rm{min}}=\frac{(2^{R_1^{\rm{min}}}-1)\sigma^2}{|h_1|^2}$. Also, we replace $P_1$ with $P_2$ by considering that equality is achieved at the minimum rate of user 2. Thus, we have $P_1 = \frac{P_2|h_2|^2}{(2^{R_2^{\rm{min}}}-1)|h_1|^2 }- \frac{\sigma^2}{|h_1|^2}= kP_2+ b $, with $k=\frac{|h_2|^2}{(2^{R_2^{\rm{min}}}-1)|h_1|^2 }$ and $b=-\frac{\sigma^2}{|h_1|^2}$. Then, the multi-variable function of the EE becomes a single variable function over $P_2$, which is given by 
\begin{equation} \label{eq:single variable}
f(P_2)=\frac{\log_2 \left( 
1+ \frac{ (kP_2+ b)|h_{1}|^2+P_2 |h_{2}|^2 }
{\sigma^2}
\right)}{P_f + kP_2+P_2+b}.
\end{equation}

Correspondingly, the root of the derivative is denoted as $\overline{P}_2^{\star}\leftarrow f^{'}(P_2)=0$. The corresponding value for $P_1$ is $\overline{P}_1^{\star}=k\overline{P}_2^{\star}+b$.

\begin{IEEEproof}
Refer to Appendix II.
\end{IEEEproof}


\begin{remark}
It can be seen that changing the SIC order leads to different PA results. Under Case II, even for Phases I and II, user 1 may not transmit at full power. For Phase I, instead, user 2 transmits at full power. Also, it is quite difficult to judge which decoding order is better, since this depends on the transmit power constraint and the \ac{QoS} requirement. {\color{black}If both constraints are the same for both users, under Phase I, it is clear that Case I always outperforms Case II. However, even in this case, except from Phase I}, it is still difficult to compare them analytically.  
\end{remark}

\begin{figure*}
\centering
\begin{subfigure}{0.5\textwidth}
  \centering
  \includegraphics[width=1\linewidth]{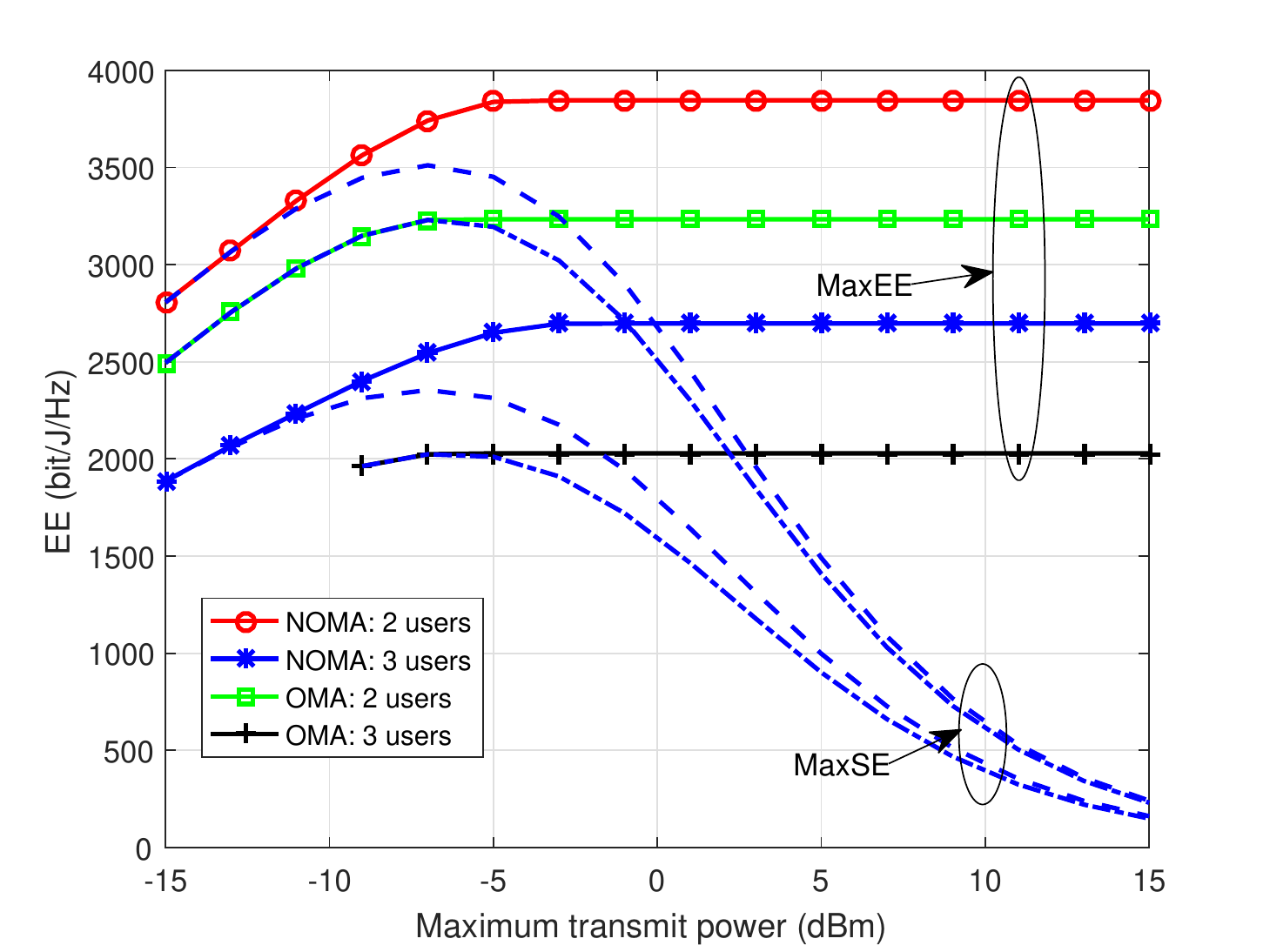}
  \caption{}
  \label{fig:sub1}
\end{subfigure}%
\begin{subfigure}{0.5\textwidth}
  \centering
  \includegraphics[width=1\linewidth]{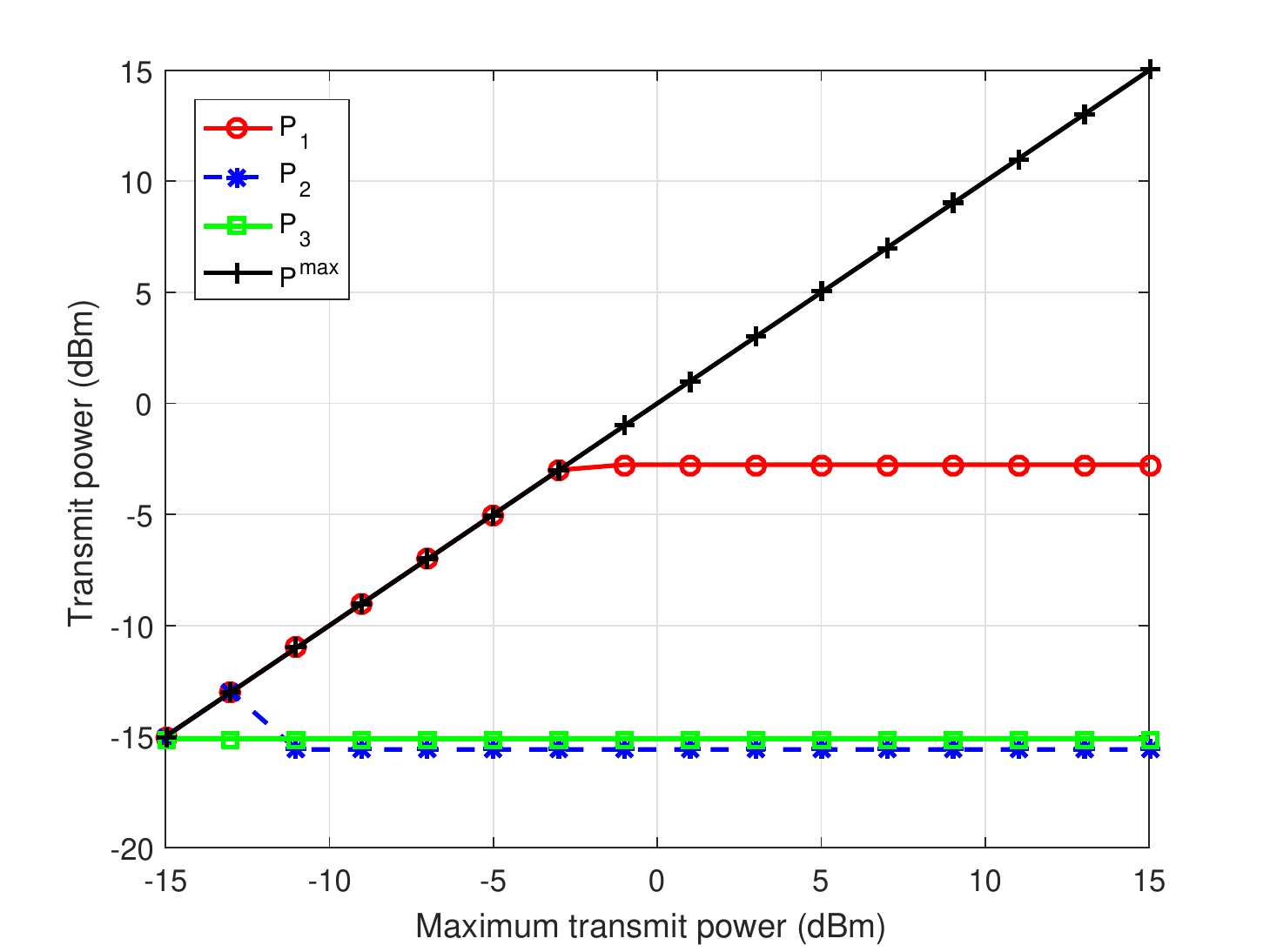}
  \caption{}
  \label{fig:sub2}
\end{subfigure}
\caption{Case I: larger channel gain difference; a) EE versus maximum transmit power; b) corresponding transmit power for three users; $|h_1|^2=1.10 \times 10^{-9}, |h_2|^2=1.34 \times 10^{-10}, |h_3|^2=4.25 \times 10^{-11}$. }
\label{fig:test}
\end{figure*}

\begin{figure*}
\centering
\begin{subfigure}{0.5\textwidth}
  \centering
  \includegraphics[width=1\linewidth]{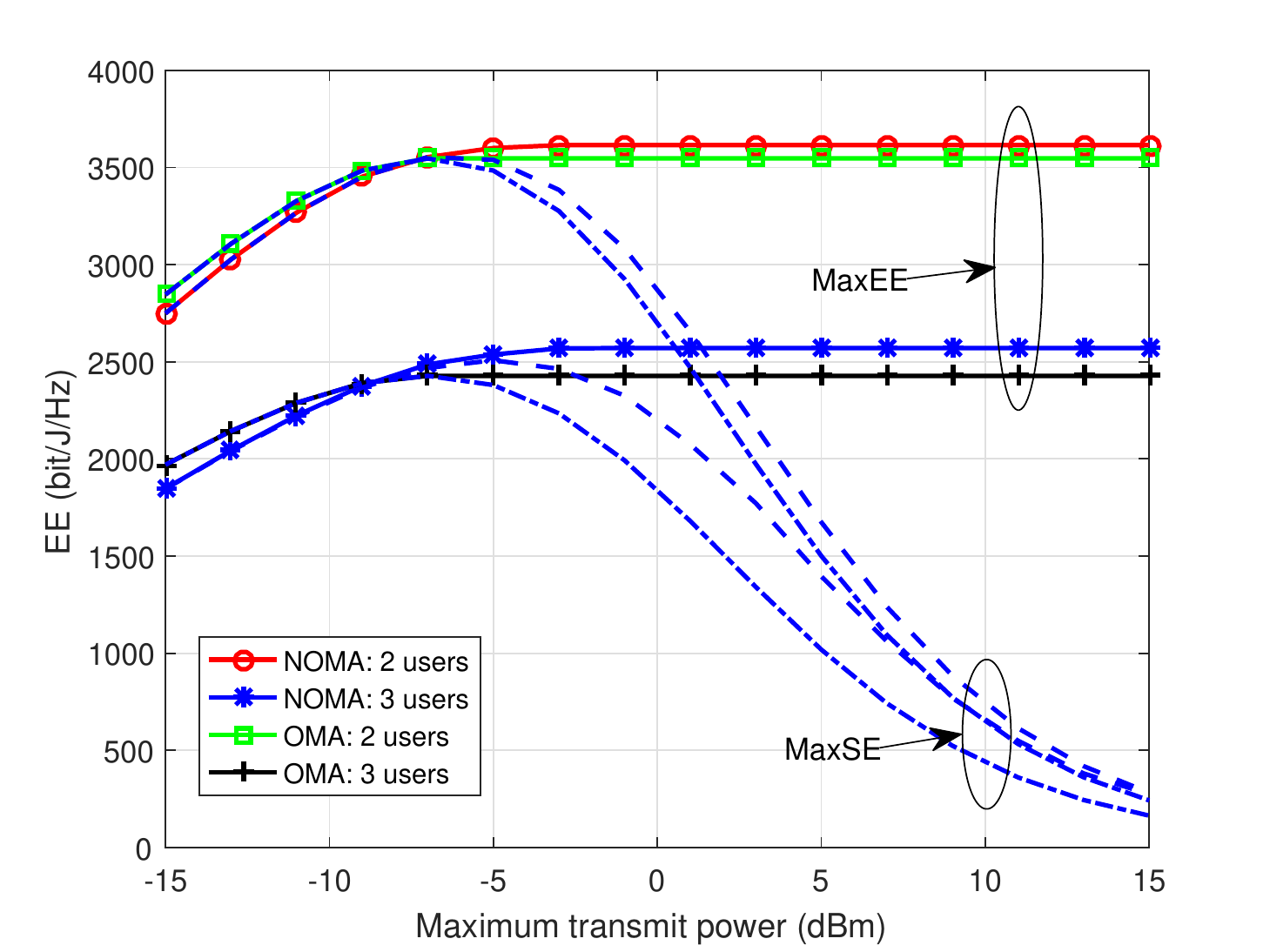}
  \caption{}
  \label{fig:sub1}
\end{subfigure}%
\begin{subfigure}{0.5\textwidth}
  \centering
  \includegraphics[width=1\linewidth]{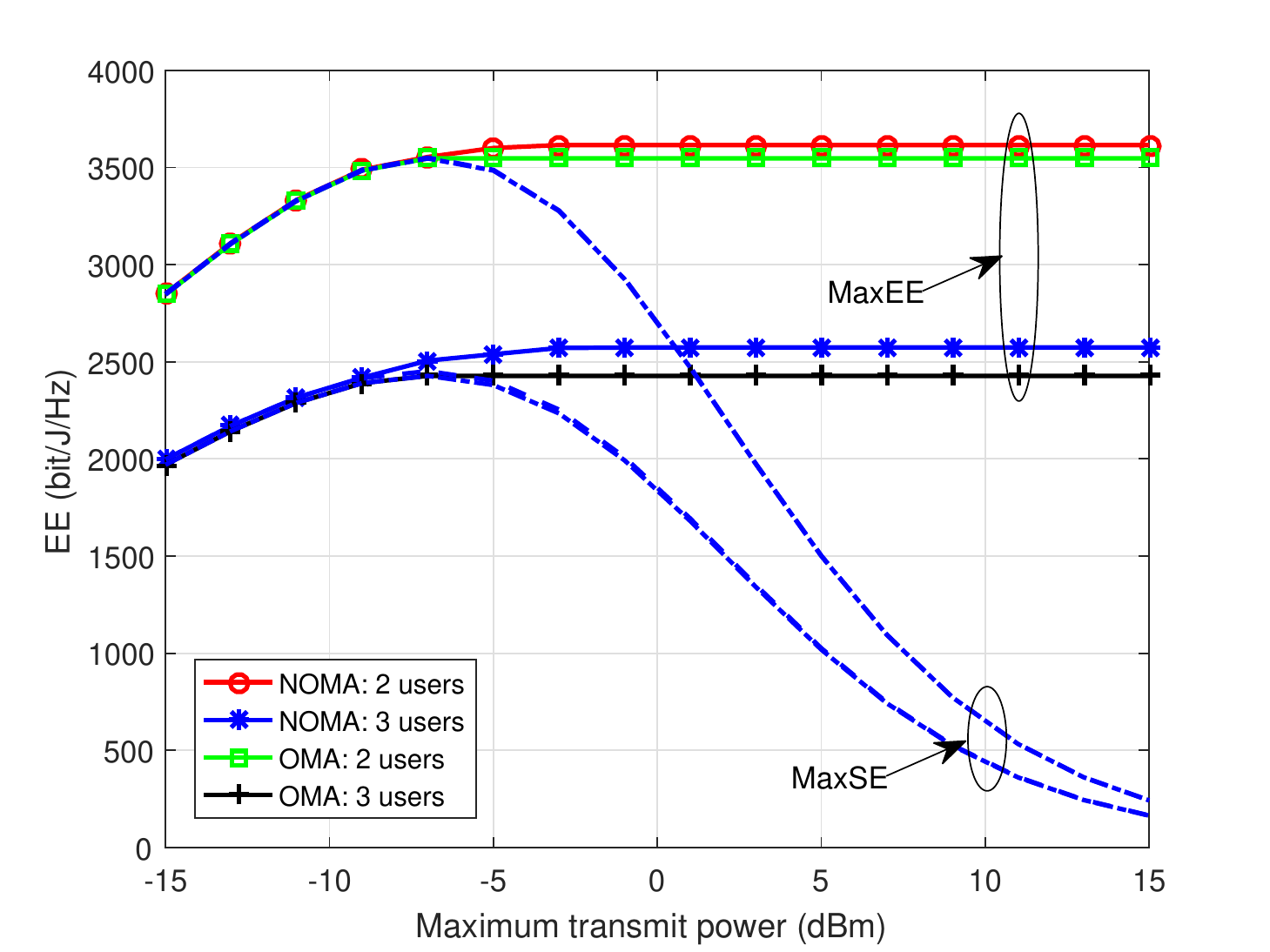}
  \caption{}
  \label{fig:sub2}
\end{subfigure}
\caption{Case II: smaller channel gain difference; a) EE versus maximum transmit power with QoS constraints; b) EE versus maximum transmit power without QoS constraints; $|h_1|^2=7.31 \times 10^{-10}, |h_2|^2=5.81 \times 10^{-10}, |h_3|^2=3.10 \times 10^{-10}$.}
\label{fig:test}
\end{figure*}

\begin{table} [!h]
\caption{Simulation Parameters.} 
\renewcommand{\arraystretch}{0.75}
\label{Table III} 
\centering
 \begin{tabular}{c|c} 
 \hline  
\bfseries Parameters & \bfseries Value \\ [0.5ex] 
 \hline\hline
 Number of users per cluster &$L=2, 3$ \\
 \hline
 Number of RBs &$M=1, 4,8$ \\
 \hline
 Minimum rate requirement &$R^{\rm{min}}=1.5$  [bit/s/Hz] \\
 \hline
 Fixed Transmit power per user & $P_f = 0$ [dBm] \\
 \hline
 Channel bandwidth per RB & $180$ [KHz] \\ 
 \hline
{Noise power spectral} density & $-174$ [dBm/Hz] \\
 \hline
 Path-loss model & $128+35\log_{10}(d)$, $d$ in kilometer \\
 \hline  
 Small scale fading & $\mathcal{CN}(0,1)$ \\
 \hline
\end{tabular}
\end{table}

\section{Simulation Results}
In this section, simulations are conducted to verify the developed results. The {\color{black}default} simulation parameters are listed in Table III.
Note that in simulations, we assign the same minimum rate requirements and maximum transmit power constraints to all users. 

\subsection{Single cluster}
Results for two cases with different channel gain difference between the users are shown in Figs. 1 and 2. In addition, as a baseline algorithm, OMA with equal degrees of freedom is presented. The results are also obtained by running the proposed Algorithm 2 with the rate expressions adjusted {according to the OMA protocol}. Moreover, we also present the results when the objective is to maximize the SE of the system, which is denoted as ``MaxSE". In contrast, the EE maximizing results are denoted as ``MaxEE". From Figs. 1(a) and 2(a), we can see that the EE first increases with the maximum transmit power for both ``MaxEE" and ``MaxSE". Then, after a certain threshold, ``MaxEE" saturates, while ``MaxSE" continues to decrease. This illustrates the importance of applying an energy-efficient PA algorithm, especially under high maximum transmit power. 

Specifically, Fig. 1 shows the case with larger channel gain difference among the users. In this case, NOMA achieves much higher EE than OMA for both two and three users, respectively. Moreover, for both NOMA and OMA, the two user case is much better than the three user case. Fig. 1(b) shows how the three users allocate their power as the maximum transmit power increases. For user 1, under low maximum transmit power, full power is consumed, which agrees with Theorem 2. Under high maximum transmit power, its power no longer increases with the maximum transmit power, but saturates. This coincides with Theorem 3. Moreover, for users 2 and 3, they are transmitting using the minimum required power, as expected based on Theorem 3. 
\begin{figure*}
\renewcommand\thefigure{4}
\centering
\begin{subfigure}{0.5\textwidth}
  \centering
  \includegraphics[width=1\linewidth]{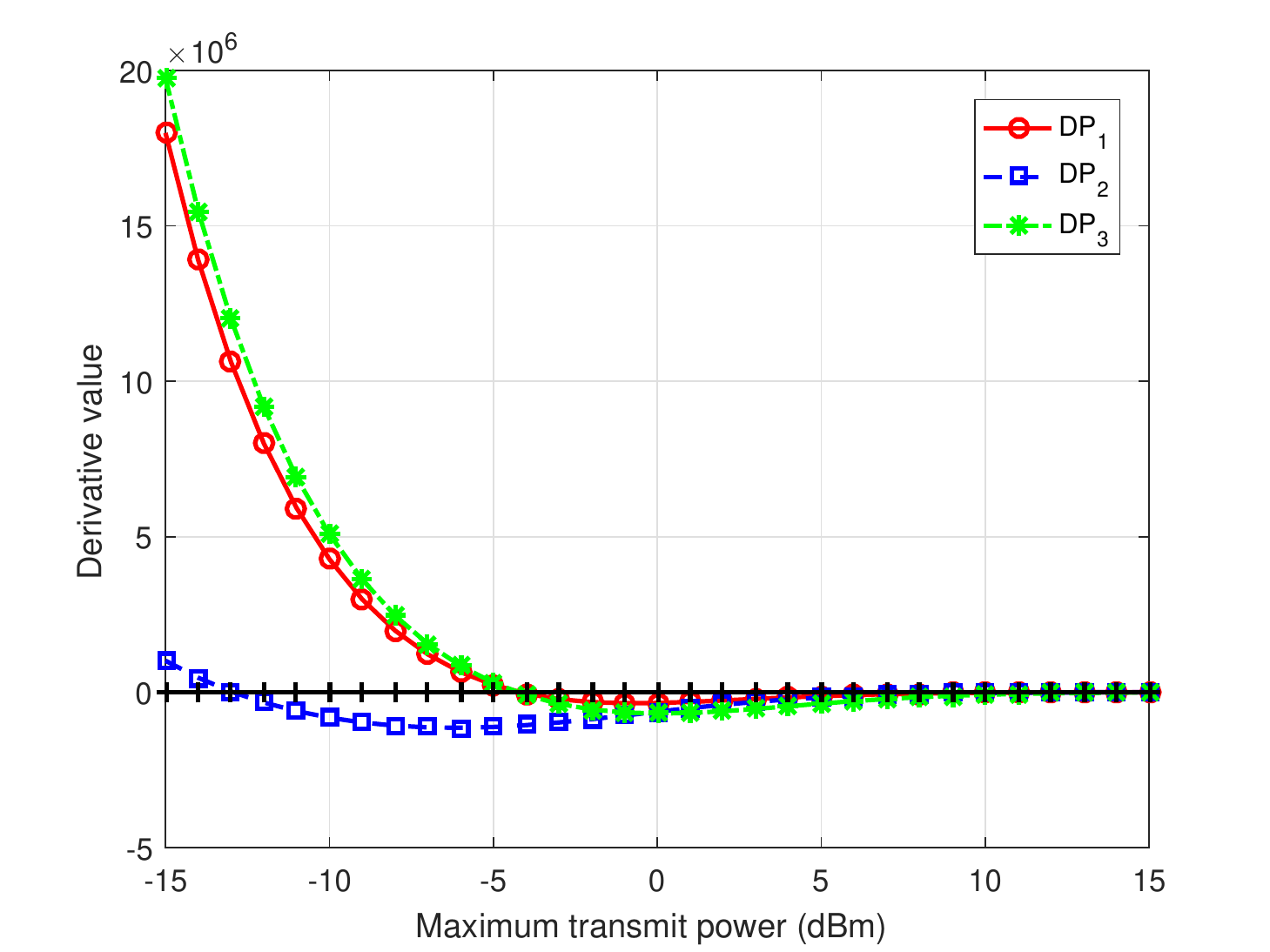}
  \caption{}
  \label{fig:sub1}
\end{subfigure}%
\begin{subfigure}{0.5\textwidth}
  \centering
  \includegraphics[width=1\linewidth]{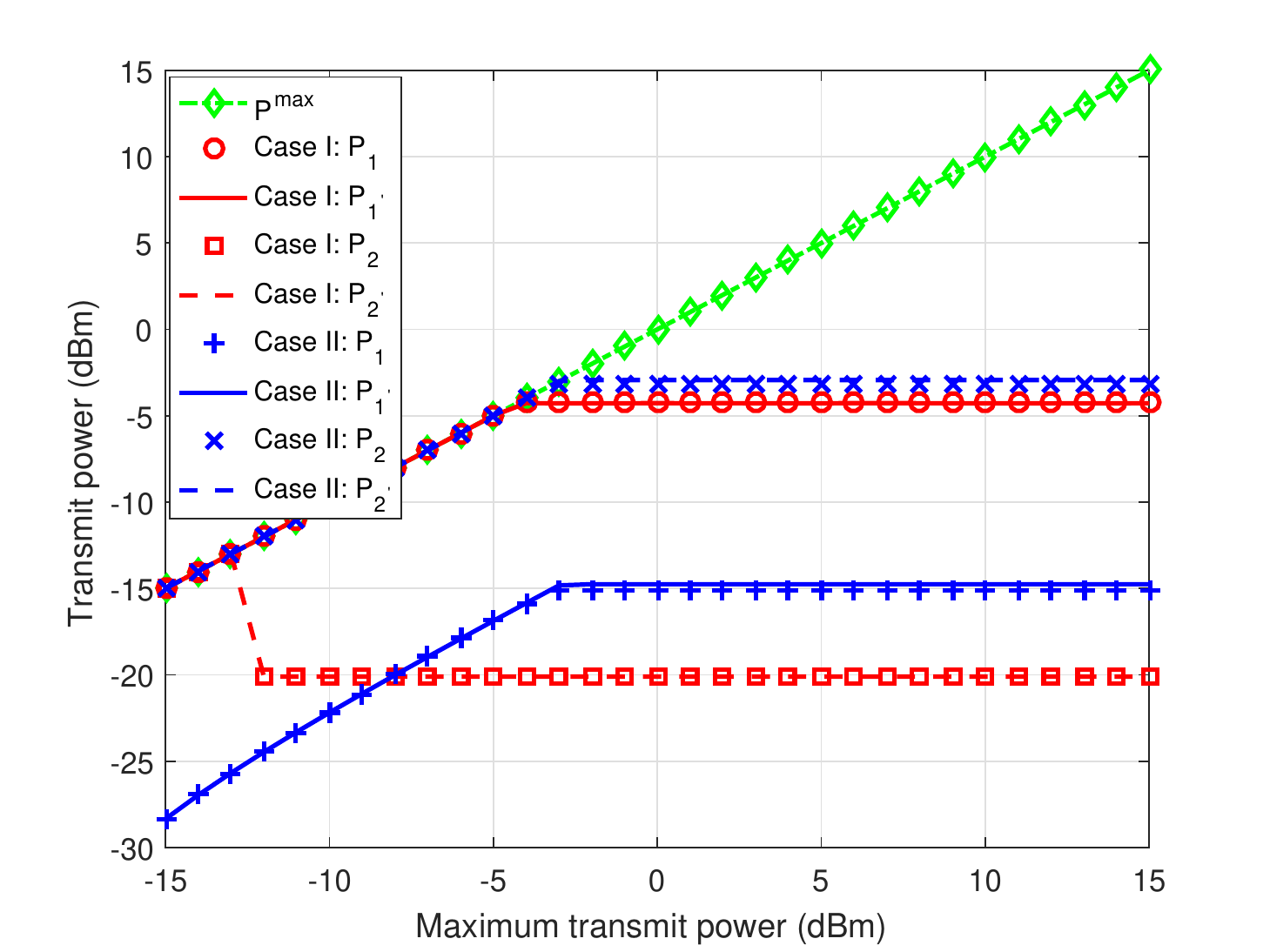}
  \caption{}
  \label{fig:sub2}
\end{subfigure}
\caption{Comparison between two SIC orders; a) Partial derivative values; b) PA; $|h_1|^2=1.10 \times 10^{-9}, |h_2|^2=1.34 \times 10^{-10}$.}
\label{fig:4}
\end{figure*}

\begin{figure}
\renewcommand\thefigure{3}
\centering
\includegraphics[width=0.5\textwidth]{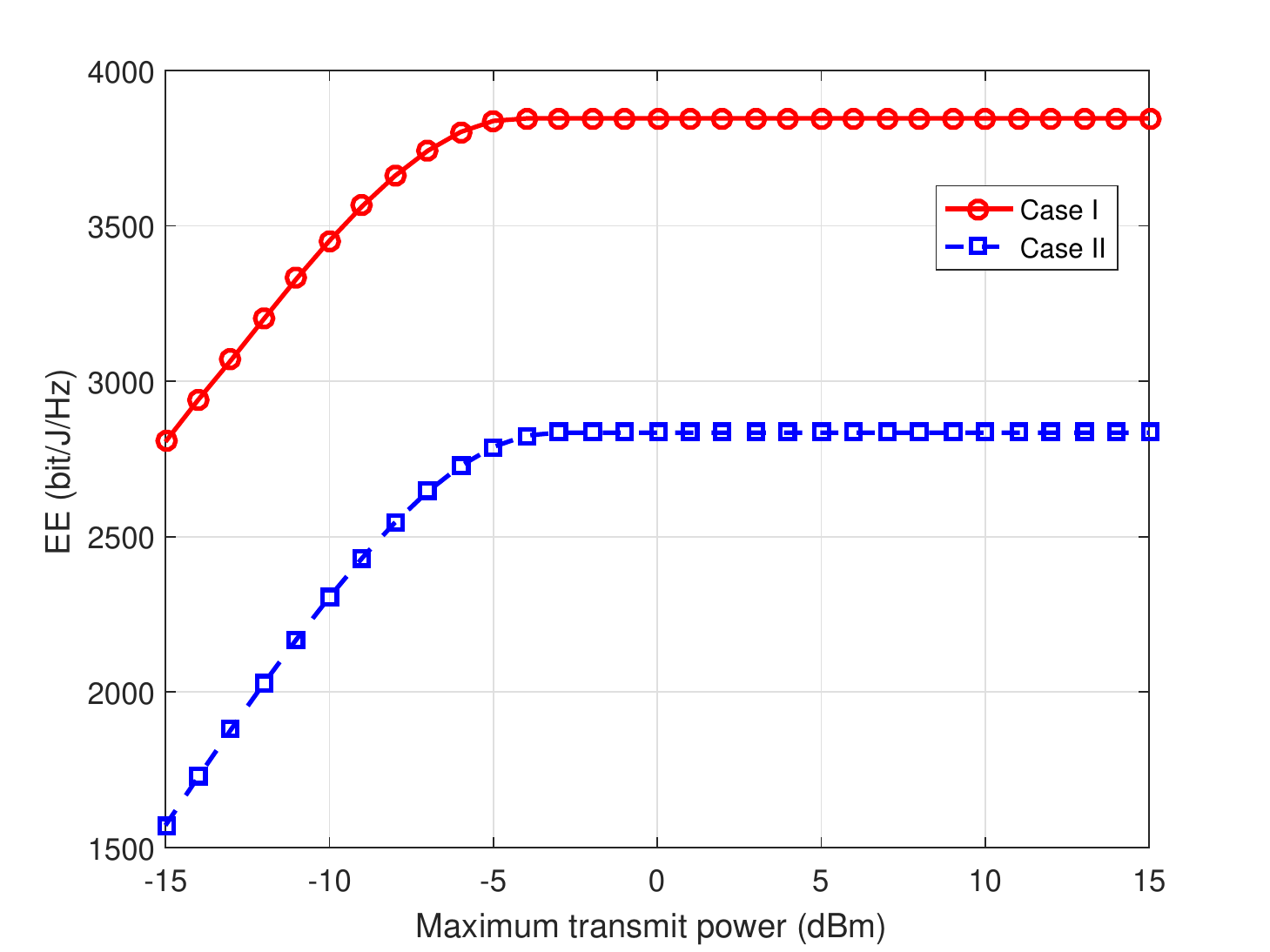}
\caption{Comparison of the EE between the two SIC orders; $|h_1|^2=1.10 \times 10^{-9}, |h_2|^2=1.34 \times 10^{-10}$.}
\label{fig:3}
\end{figure}

Fig. 2 shows the case of smaller channel gain difference. In this case, for both OMA and NOMA, the two user case is better than the three user case. However, under low maximum transmit power, OMA is better than NOMA. This is because the interference introduced by NOMA leads to a smaller feasibility region. Take two user case for example, under low maximum transmit power, OMA is transmitting at full power for both users. However, if $ \frac{P_1^{\rm{max}}|h_1|^2}{(2^{R_1^{\rm{min}}}-1)|h_2|^2 }- \frac{\sigma^2}{|h_2|^2} <P_2^{\rm{max}}$, \text{user 2} in NOMA cannot transmit at full power to ensure the QoS of user 1. This is quite different from the downlink case, in which the BS controls the PA for all users, and can distribute power among them. In uplink, each user is constrained by its own maximum transmit power. Since user 1's power cannot be increased over its maximum transmit power, the allowable power for user 2 cannot be increased either. Consequently, NOMA achieves lower EE than OMA. In Fig. 2(b), we show the case when there is no \ac{QoS} constraints. As expected, NOMA is always better than OMA, even though the gain is quite minor for the two user case. Comparing this with Fig. 1, it implies that user pairing should be conducted such that the users' channel gain should be distinct. Moreover, it is worth mentioning that NOMA still outperforms OMA under high maximum transmit power.

In Figs. 3 and 4, we compare the performance between the two SIC orders. According to Fig. 3, Case I achieves much higher EE than Case II. Fig. 4(a) shows how the partial derivative values vary with the maximum transmit power, where $\text{DP}_1$, $\text{DP}_2$ and $\text{DP}_3$ denote $\frac{\partial \eta_{\rm{EE}}}{\partial P_1}|_{P_1^{\rm{max}},P_2^{\rm{max}}}, ~\frac{\partial \eta_{\rm{EE}}}{\partial P_2}|_{P_1^{\rm{max}},P_2^{\rm{max}}}$ and $ \frac{\partial \eta_{\rm{EE}}}{\partial P_2}|_{P_1^{\rm{max}},P_2^{\rm{min}}}$, respectively. It can be seen that as the maximum transmit power increases, Case I moves from Phase I to Phase IV, while Case II moves from Phase I to Phase III. 
In Fig. 4(b), we plot $P_1$ and $P_2$ obtained by Algorithm 2 (without prime) and the proposed analytical solution (with prime). Obviously, the same results are obtained by both methods, which demonstrates the correctness of the analytical solution.\footnote{As we use the log value on the y-axis, it may seem that for Case II, there exists a difference between these two algorithms. Indeed, the difference is smaller than $10^{-5}$, and it exists simply because the root can only be approximated using the bisection method.} Particularly, under low maximum transmit power, the system is in Phase I, user 1 in Case I and user 2 in Case II transmit at full power. 


%

\subsection{Multiple clusters}
In this subsection, multiple clusters are considered. 
{\color{black}We compare the proposed solution, denoted as HMA-prop, with two OMA-based algorithms, i.e., OMA-swap and OMA-MWM, and three HMA-based algorithms, i.e., HMA-DC \cite{20}, HMA-MWM and HMA-rand. OMA-swap follows the same procedure as HMA-prop, but with the rate expressions adjusted according to the OMA protocol. In OMA-MWM, we update the PA and user-RB association alternately until convergence. More exactly, under a given PA, it is clear that the EE maximization is equivalent to the sum rate maximization. According to the OMA protocol, the achievable rate of user $(m,l)$ is $R_{m,l}^{\rm{O}}=\frac{1}{L_m} \log_2 \left(1+\frac{L_m P_{m,l} |h_{m,l}|^2 }{ \sigma^2 } \right)$, which depends only on the allocated RB.
Consider the users and RBs as the two set of nodes in a bipartite graph, and the corresponding rates $R_{m,l}^{\rm{O}}$ as the weights. Then, the matching that maximizes the sum weight also maximizes the sum rate, and further the EE. This matching can be obtained efficiently using standard maximum weight matching (MWM) algorithms, such as the Hungarian algorithm \cite{Hungarian}. Under a given user-RB association, the PA can be solved using Algorithm 2, with the rate expressions adjusted according to the OMA protocol. Note that convergence is guaranteed since the EE increases or remains unchanged after each update, and there exists an upper bound. 

\begin{figure*}
\centering
\begin{subfigure}{0.5\textwidth}
  \centering
  \includegraphics[width=1\linewidth]{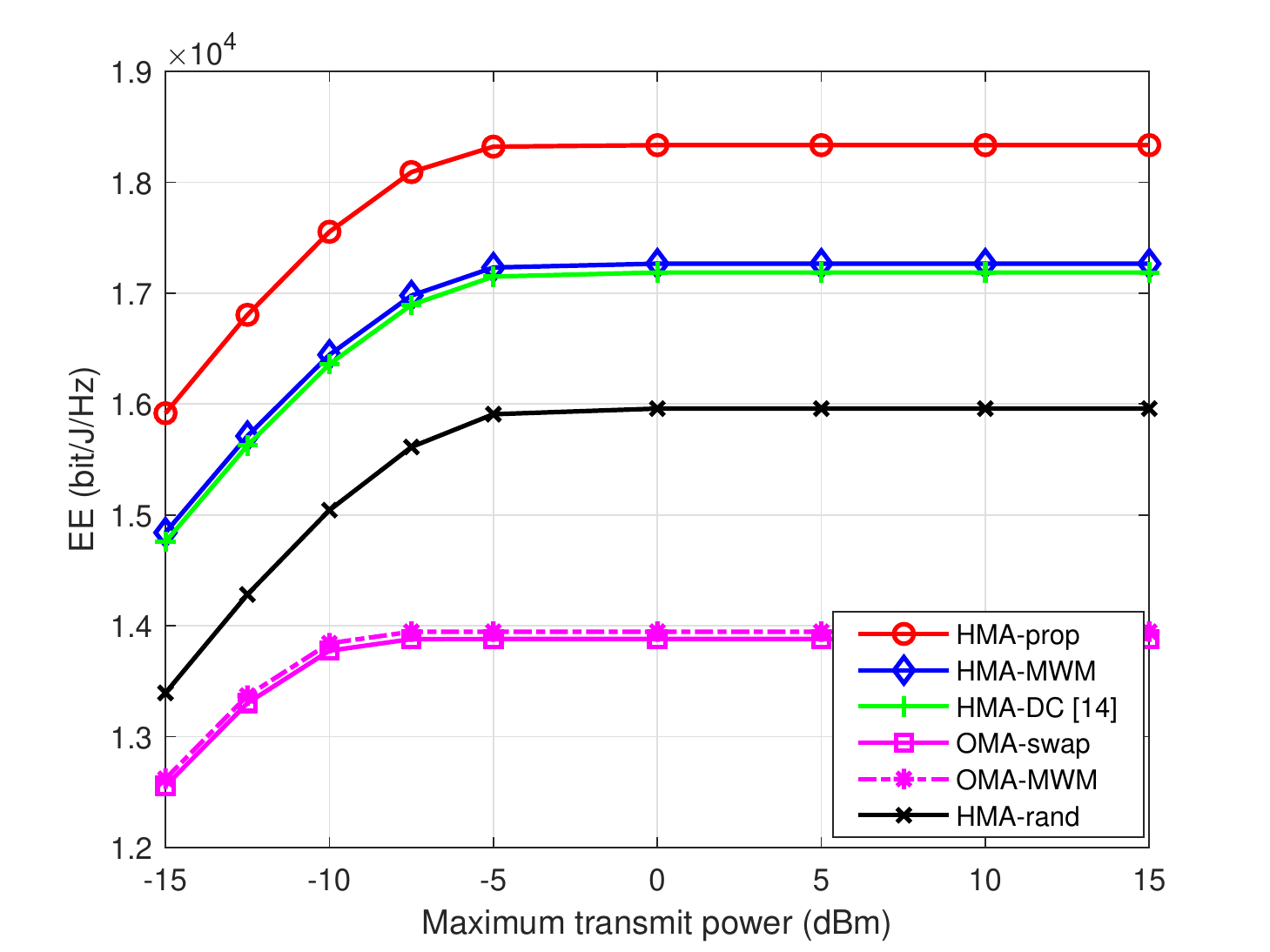}
  \caption{}
  \label{fig:sub1}
\end{subfigure}%
\begin{subfigure}{0.5\textwidth}
  \centering
  \includegraphics[width=1\linewidth]{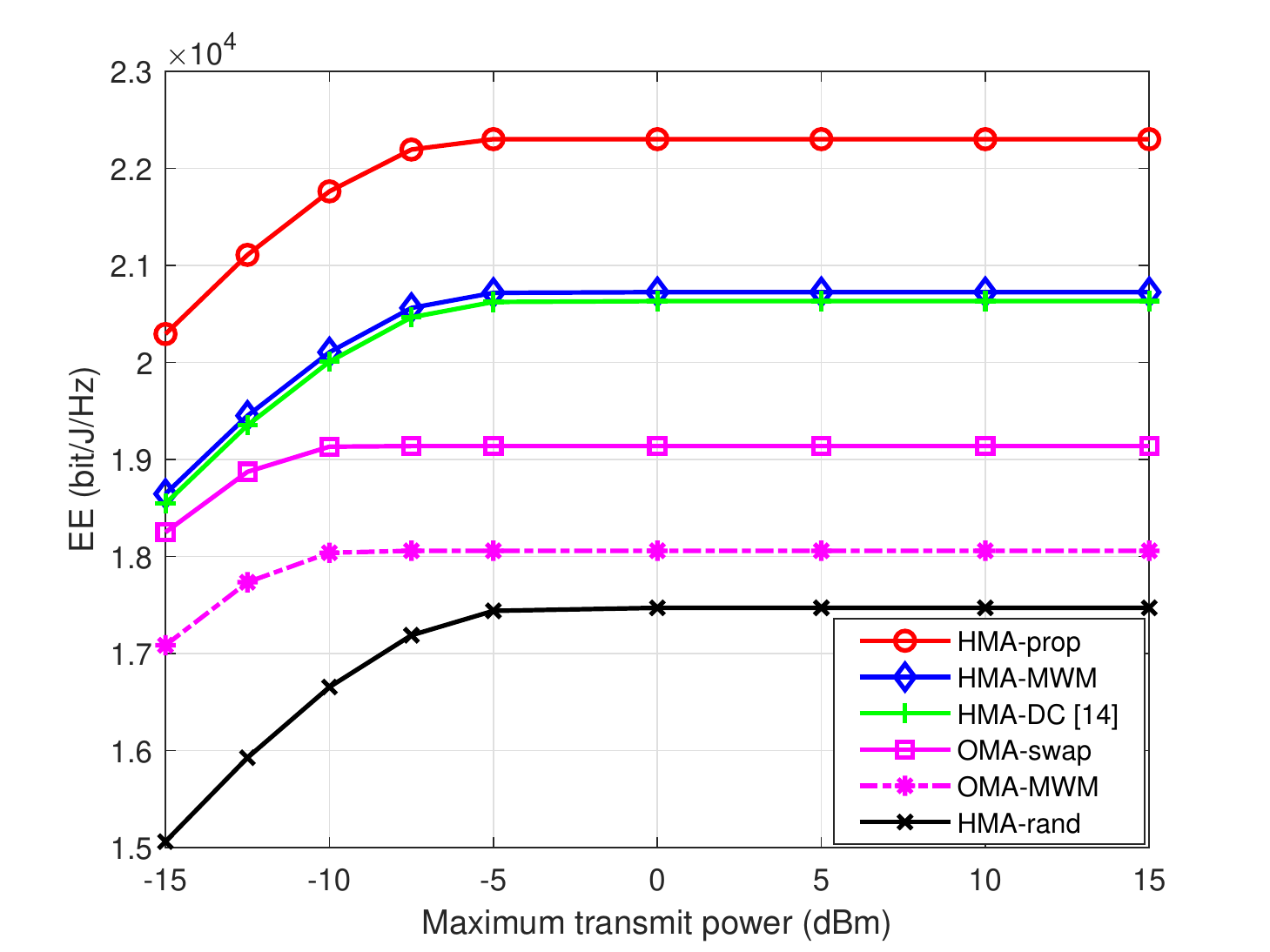}
  \caption{}
  \label{fig:sub2}
\end{subfigure}
\caption{{\color{black}Comparison of average EE when $U=12$ and $M=4$; a) smaller channel gain difference; b) larger channel gain difference.}}
\label{fig:test}
\end{figure*}

\begin{figure*}
\centering
\begin{subfigure}{0.5\textwidth}
  \centering
  \includegraphics[width=1\linewidth]{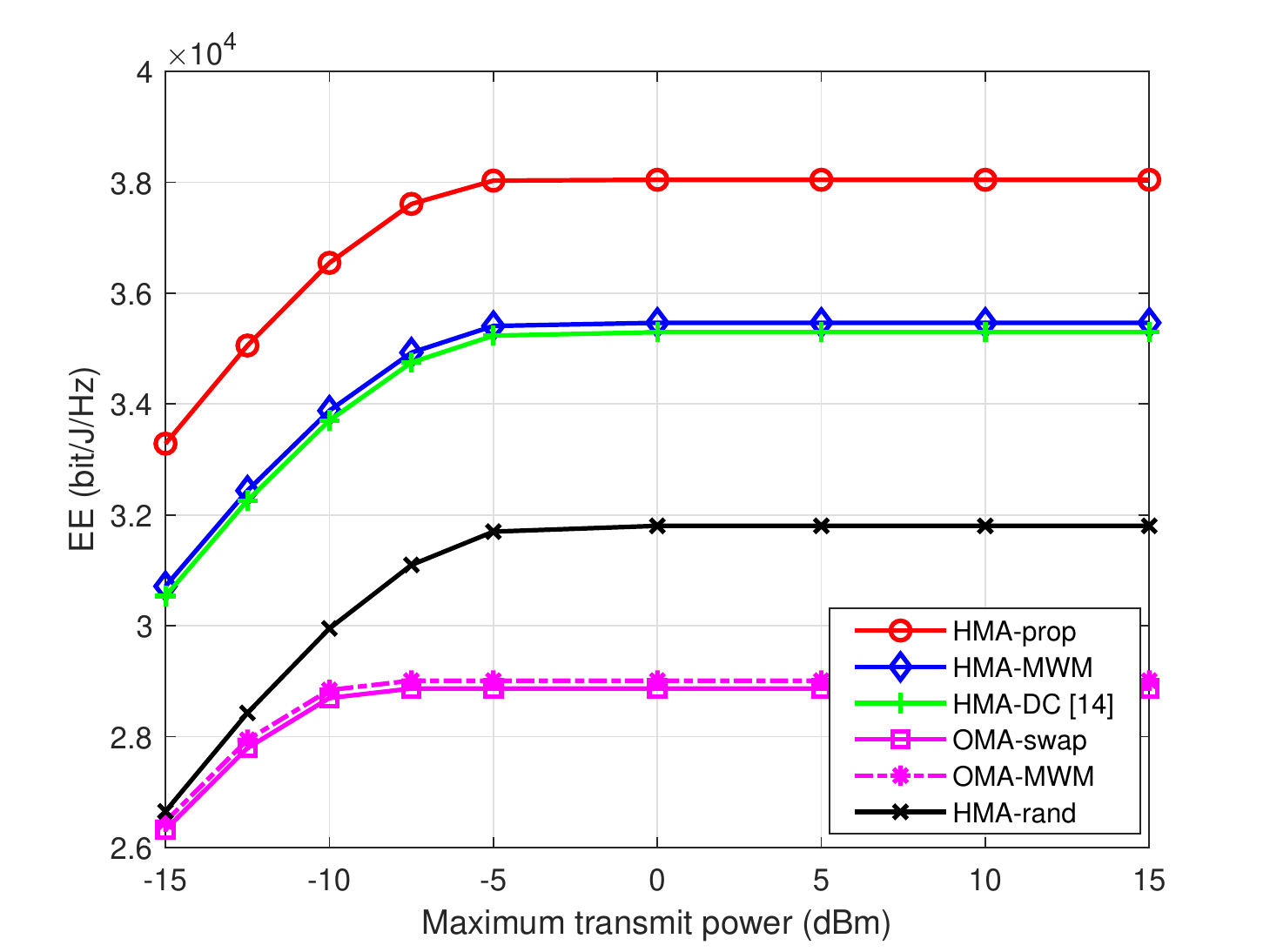}
  \caption{}
  \label{fig:sub1}
\end{subfigure}%
\begin{subfigure}{0.5\textwidth}
  \centering
  \includegraphics[width=1\linewidth]{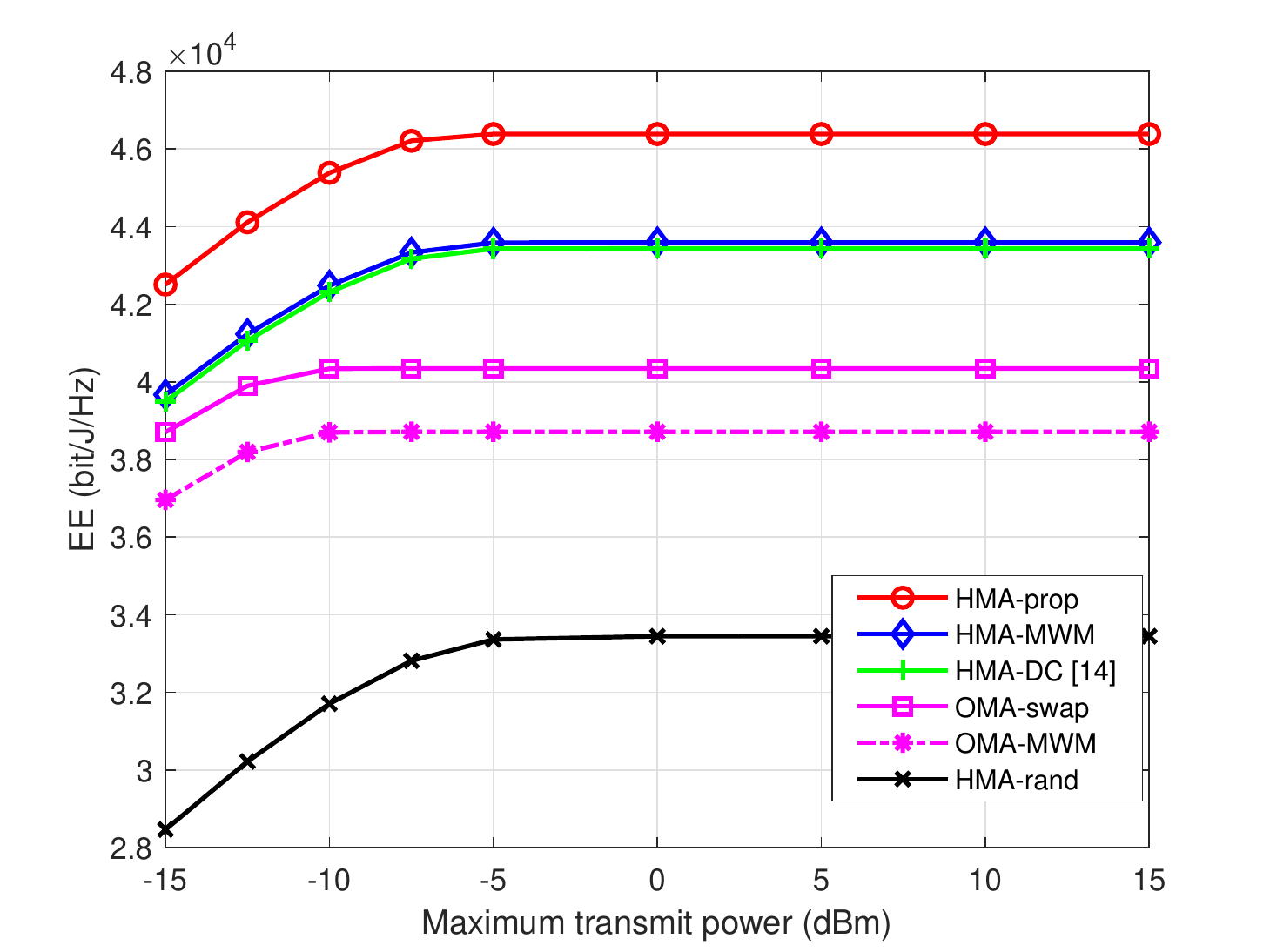}
  \caption{}
  \label{fig:sub2}
\end{subfigure}
\caption{{\color{black}Comparison of average EE when $U=24$ and $M=8$; a) smaller channel gain difference; b) larger channel gain difference.}}
\label{fig:test}
\end{figure*}

HMA-DC is the scheme proposed in \cite{20}, in which each user sends its matching request to its most preferred RB based on the channel gain. However, the preferred RB only accepts the users that lead to the maximum EE. The rejected users will move to the next preferred RB and this process continues until all users are matched to an RB. For HMA-MWM, we cannot apply it the same way as for OMA-MWM, since MWM for HMA cannot be performed due to the intra-cluster interference. Instead, we simply consider the weights to be the channel gains. Then, we conduct the MWM between the users and the RBs to achieve the maximum sum channel gains.
In HMA-rand, the users are allocated to the RBs randomly. 

The following results are averaged over $10^3$ random trials, and for each trial, we generate the users' locations following a uniform distribution. We first present the result when there are 12 users accessing 4 RBs. We consider two cases with different channel gain differences. Fig. 5(a) shows the result when all users lie within a radius of 150 m. In Fig. 5(b), the users are equally divided into three circles, with radii of 50, 100 and 150 m, respectively. Therefore, \text{Fig. 5(a)} is the case with smaller channel gain difference. It is clear that HMA-prop is the best, followed by HMA-MWM, HMA-DC, HMA-rand OMA-MWM and OMA-swap. This validates the superiority of the proposed scheme over other HMA- and OMA-based algorithms. In Fig. 5(b), it can be seen that HMA-prop is still the best, followed by HMA-MWM and HMA-DC. However, in this case, OMA-swap outperforms OMA-MWM, and HMA-rand is the worst. Quite surprisingly, by comparing Figs. 5(a) and 5(b), we can observe that a larger channel gain difference does not necessarily lead to a larger gain of HMA over OMA. This does not contradict the conclusion in the single cluster, where we claim that a large channel gain difference among users yields a larger gain of HMA over OMA. This is because the user-RB association results in HMA and OMA can be quite different, and the former conclusion holds when the user-RB association remains the same for both schemes. 

Figure 6 shows the result when we double the number of users and RBs, i.e., now there are 24 users accessing 8 RBs. By comparing Figs. 5 and 6, it is clear that the corresponding EE values in Fig. 6 are more than twice those in Fig. 5, except for HMA-rand. This implies that a multiplexing gain is obtained by having more RBs.  

Figure 7 plots the cumulative distribution function (CDF) of the number of swapping operation required to reach convergence for the above two scenarios. It can be seen that the number of swapping operations grows with that of RBs. However, less than 60 swapping operations are needed even for the scenario when $U=24$ and $M=8$, which is quite small. In addition, for Algorithm 3, simulation results show that exactly 6 iterations are required for it to converge when there are 3 users sharing an RB.

}

\begin{figure}
\centering
\includegraphics[width=0.5\textwidth]{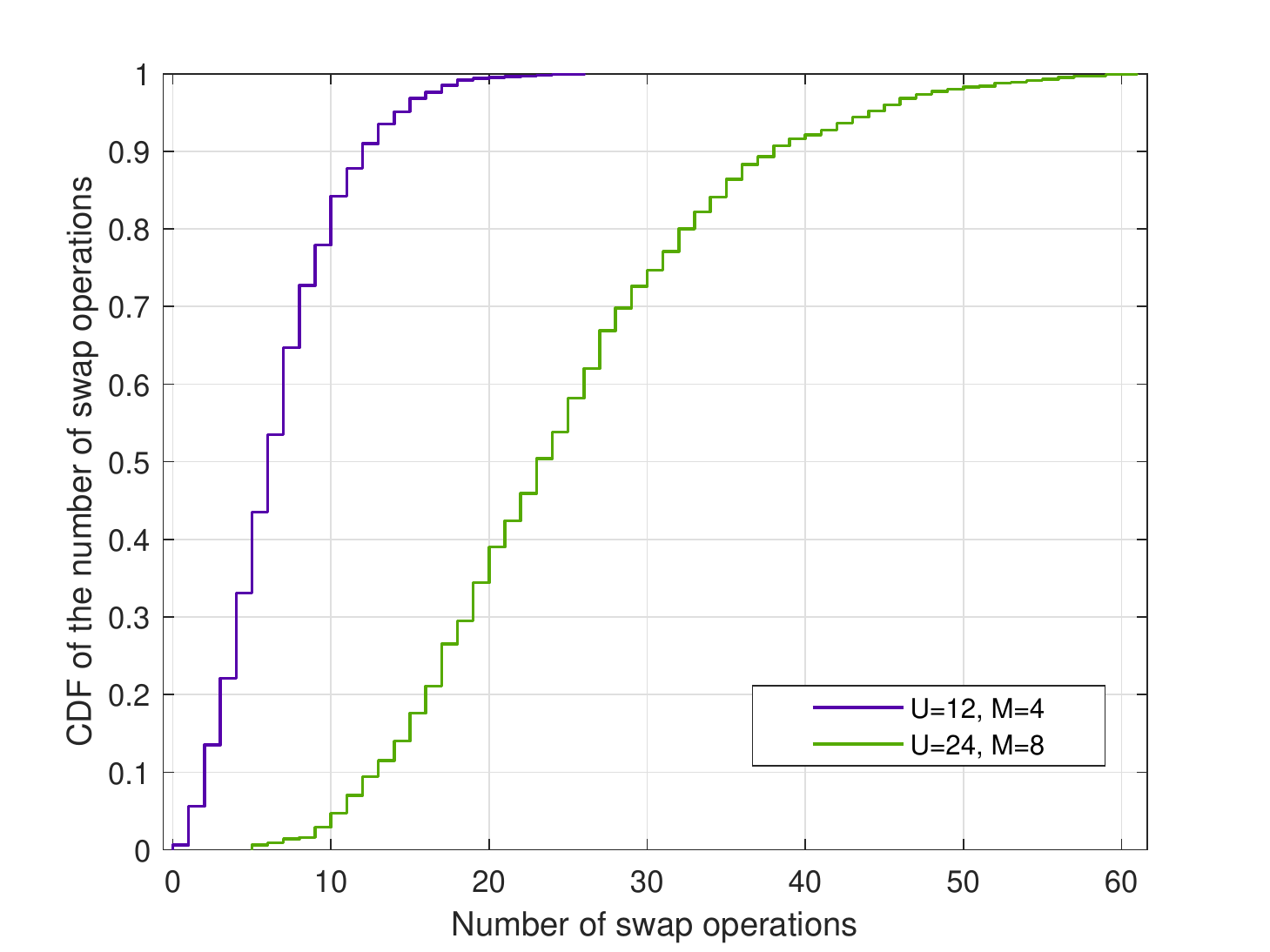}
\caption{CDF of the number of swap operations for convergence.}
\end{figure}

\section{Conclusion}
In this paper, we have studied the energy-efficient resource allocation for HMA uplink with QoS requirements for each user. {\color{black}Based on swap matching in many-to-one bipartite graph, we have proposed a joint user-RB association and power allocation scheme, which is guaranteed to converge. Under a given user-RB association, we have shown that the system EE maximization equals cluster EE maximization.  Then, we have derived the feasibility conditions, and proposed to solve the EE maximization using Dinkelbach's algorithm. Moreover, to further relieve the computational burden, a low-complexity optimal algorithm has been proposed for solving the convex optimization subproblem inside the Dinkelbach's algorithm. For the two user case, analytical solutions have further been derived for the two SIC orders.
Simulations have been performed, which verify the developed analytical solutions. Moreover, the results for a single cluster show that under low maximum transmit power, OMA can be better than HMA for uplink, due to the smaller feasibility region for HMA caused by the QoS requirements. On the other hand, under high maximum transmit power, HMA still outperforms OMA. Results under multiple clusters fully validate the superiority of the proposed scheme over other HMA- and OMA-based algorithms. Furthermore, a multiplexing gain can be observed when employing more RBs.}  

{\color{black}In this work, we have considered the situation in which each user can access a single RB. The extension to multiple RBs is interesting for future work.}

\appendices
\section{Proof of Table I}
In Phases I, II and III, as it satisfies the condition for Theorem 2, we can conclude that $P_1 = P_1^{\rm{max}}$. As for Phase IV, it is exactly the condition for Theorem 3, and thus, the conclusion holds. Then, we only need to prove the PA for user 2 in Phases I, II and III.  

Let us first consider Phase I. For the differentiable strictly pseudo-concave function, since $\frac{\partial \eta_{\rm{EE}}}{\partial P_1}|_{P_1^{\rm{max}},P_2^{\rm{max}}} \geq \frac{\partial \eta_{\rm{EE}}}{\partial P_2}|_{P_1^{\rm{max}},P_2^{\rm{max}}} \geq 0$, we can conclude that $\frac{\partial \eta_{\rm{EE}}}{\partial P_1} \geq \frac{\partial \eta_{\rm{EE}}}{\partial P_2} \geq 0$ for any value of $P_1$ and $P_2$. Thus, increasing the transmit power for each user leads to an larger EE. However, for \text{user 2}, increasing $P_2$ also causes more interference to user 1. To ensure the \ac{QoS} requirement of \text{user 1}, the maximum power can be used by user 2 is given by $\frac{P_1^{\rm{max}}|h_1|^2}{(2^{R_1^{\rm{min}}}-1)|h_2|^2 }- \frac{\sigma^2}{|h_2|^2}$. Combining this with the transmit power constraint, we have $P_2 = \min \left(P_2^{\rm{max}}, \frac{P_1^{\rm{max}}|h_1|^2}{(2^{R_1^{\rm{min}}}-1)|h_2|^2 }- \frac{\sigma^2}{|h_2|^2} \right)$.

Next, let us focus on Phases II and III. Without considering the \ac{QoS} constraint, $P_2$ is obtained when $\frac{\partial \eta_{\rm{EE}}}{\partial P_2}|_{P_1^{\rm{max}}}=0$. Denote it as $P_2^{\star}$, satisfying $P_2^{\star} < P_2^{\rm{max}}$. On the other hand, due to the minimum rate requirements for user 1 and user 2, $P_2$ has a lower bound $P_2^{\rm{min}}$, and an upper bound $\frac{P_1^{\rm{max}}|h_1|^2}{(2^{R_1^{\rm{min}}}-1)|h_2|^2 }- \frac{\sigma^2}{|h_2|^2} $. If $P_2^{\star} \leq P_2^{\rm{min}}$, $P_2 = P_2^{\rm{min}}$. Otherwise, $P_2 = \min \left( P_2^{\star}, \frac{P_1^{\rm{max}}|h_1|^2}{(2^{R_1^{\rm{min}}}-1)|h_2|^2 }- \frac{\sigma^2}{|h_2|^2} \right)$.

\section{Proof of Table II}
In Phase I, due to the change of SIC order, user 2 should transmit at full power. As for user 1, it should not violate the QoS requirement of user 2, and thus, $P_1 < \frac{P_2^{\rm{max}}|h_2|^2}{(2^{R_2^{\rm{min}}}-1)|h_1|^2 }- \frac{\sigma^2}{|h_1|^2}$. Combining it with the maximum power constraint, we have $P_1 = \min \left(P_1^{\rm{max}}, \frac{P_2^{\rm{max}}|h_2|^2}{(2^{R_2^{\rm{min}}}-1)|h_1|^2 }- \frac{\sigma^2}{|h_1|^2} \right)$. 

In Phases II and III, we first assume that $P_1$ is constrained by the QoS of user 2. Then, we turn the multi-variable function into a single variable function. 
Accordingly, the solution for this is the root of the derivative. Denote this as $\overline{P}_2^{\star}$. Correspondingly, $\overline{P}_1^{\star}=k\overline{P}_2^{\star}+b$. 
On the other hand, \text{user 1} needs to satisfy its own QoS, and thus, we obtain $\overline{P}_1^{\rm{min}}$. If $\overline{P}_1^{\star}<\overline{P}_1^{\rm{min}}$, $P1=\overline{P}_1^{\rm{min}}$, and $P_2= (\overline{P}_1^{\rm{min}}-b)/k$. If $P_2 < (\overline{P}_1^{\rm{min}}-b)/k$, it cannot satisfy its own QoS. If $P_2$ exceeds this, EE decreases, since $P_2>\overline{P}_2^{\star}$, and $P_1>\overline{P}_1^{\star}$. When $\overline{P}_1^{\star}$ lies in $(\overline{P}_1^{\rm{min}}, P_1^{\rm{max}})$, if $\overline{P}_2^{\star} \leq P_2^{\rm{max}}$, $P_1=\overline{P}_1^{\star},~P_2=\overline{P}_2^{\star}$ is clearly the solution. As for the case $ \overline{P}_2^{\star} > P_2^{\rm{max}}$, this cannot hold. This is because we can transfer power from user 2 to user 1, and increase the EE. Therefore, $\overline{P}_2^{\star}$ cannot be the root of \eqref{eq:single variable}.
When $\overline{P}_1^{\star} >P_1^{\rm{max}}$, $P_1=P_1^{\rm{max}}$. Denote the root of
 $ \frac{\partial \eta_{\rm{EE}}}{\partial P_2}|_{P_1^{\rm{max}}}$ as $P_2^{\rm{r}}$. Since $\frac{\partial \eta_{\rm{EE}}}{\partial P_2}|_{P_1^{\rm{max}},P_2^{\rm{max}}}\leq 0$, $P_2^{\rm{r}} \leq P_2^{\rm{max}}$. In addition, we can obtain  $P_2^{\rm{r}} \geq (P_1^{\rm{max}}-b)/k $, owing to $\frac{\partial \eta_{\rm{EE}}}{\partial P_2}|_{P_1^{\rm{max}}, (P_1^{\rm{max}}-b)/k}  \geq \frac{\partial \eta_{\rm{EE}}}{\partial P_2}|_{\overline{P}_1^{\star}, \overline{P}_2^{\star}} =0$. Therefore, the QoS of user 2 can be satisfied when $P_1=P_1^{\rm{max}}$ and $ P_2= P_2^{\rm{r}}$. In sum, we can conclude that $ P_2= P_2^{\rm{r}}$.

\bibliographystyle{IEEEtran}
\balance
\bibliography{IEEEabrv,conf_short,jour_short,mybibfile}

\end{document}